\RequirePackage{ifpdf}
\documentclass[12pt,letterpaper]{JHEP3}         
\usepackage{amssymb,amsfonts}

\usepackage{wrapfig}
\usepackage{epsfig}
\usepackage{youngtab}

\def\be{\begin{eqnarray}}
\def\ee{\end{eqnarray}}
\newcommand{\nn}{\nonumber}
\newcommand\para{\paragraph{}}

\newcommand{\eqn}[1]{(\ref{#1})}

\def\Dslash{\,\,{\raise.15ex\hbox{/}\mkern-12mu D}}
\def\Dbarslash{\,\,{\raise.15ex\hbox{/}\mkern-12mu {\bar D}}}
\def\delslash{\,\,{\raise.15ex\hbox{/}\mkern-9mu \partial}}
\def\delbarslash{\,\,{\raise.15ex\hbox{/}\mkern-9mu {\bar\partial}}}
\def\pslash{\,\,{\raise.15ex\hbox{/}\mkern-9mu p}}
\def\calDslash{\,\,{\raise.15ex\hbox{/}\mkern-12mu {\cal D}}}

\def\Icurr{{\cal J}} %
\def\lae{\mathrel{\mathop{\smash{\lower .5 ex \hbox{$\stackrel<\sim$}}}}}
\def\lae{\mathrel{\mathop{\smash{\lower .5 ex \hbox{$\stackrel>\sim$}}}}}

\title{A Matrix Model for WZW}

\author{Nick Dorey${}^1$, David Tong${}^{1,2,3}$ and Carl Turner${}^1$\\
${}^1$Department of Applied Mathematics and Theoretical Physics, \\
University of Cambridge, Cambridge, CB3 OWA, UK \\
${}^2$Department of Theoretical Physics\\ TIFR, Homi Bhabha Road, Mumbai 400 005, India \\
${}^3$Stanford Institute for Theoretical Physics\\
 Via Pueblo, Stanford, CA 94305, USA\\
{\tt  n.dorey, d.tong, c.p.turner@damtp.cam.ac.uk}\\}

\abstract{We study a $U(N)$ gauged matrix quantum mechanics which, in the large
$N$ limit, is  closely related to the chiral WZW conformal field theory. This
manifests itself in two ways. First, we construct the left-moving Kac-Moody algebra from
matrix degrees of freedom. Secondly, we compute the partition function of the
matrix model  in terms of Schur and Kostka polynomials and show that, in the
large $N$ limit, it coincides with the partition function of the  WZW model.
This same matrix model was recently shown to describe non-Abelian quantum Hall
states and the relationship to the WZW model can be understood in this
framework.}

\begin{document}
\pagestyle{plain} \setcounter{page}{1}
\newcounter{bean}
\baselineskip16pt \setcounter{section}{0}

\section{Introduction}


The purpose of this paper is to describe how a simple quantum mechanical matrix
model is related to the chiral WZW conformal field theory in $d=1+1$ dimensions.

\para

The matrix model consists of a $U(N)$ gauge field, which we denote as $\alpha$,
coupled to a complex adjoint scalar $Z$ and $p$ fundamental scalars
$\varphi_i$, $i=1,\ldots,p$. The action is first order in time derivatives and
given by
\be S  =    \int dt\ \left[ i\,{\rm tr} \left(Z^\dagger {\cal D}_t Z\right)
+ i\sum_{i=1}^p   \varphi_i^\dagger {\cal D}_t\varphi_i - (k+p)\,{\rm
tr}\,\alpha
- {\omega}\, {\rm tr}\,Z^\dagger Z \right]\label{matrix}\ee
The covariant derivatives are ${\cal D}_t Z = \partial_t{Z} - i[\alpha,Z]$ and
${\cal D}_t\varphi_i = \partial_t{\varphi}_i - i \alpha\varphi_i$ and ${\rm
tr}$ denotes the trace over $U(N)$ gauge indices. Here and in the
following $k$ is a positive integer. 
%
%

\para
In addition to the $U(N)$ gauge symmetry, the quantum mechanics has an $SU(p)$
global symmetry. We will show that, in the large $N$ limit, this matrix model
captures the physics of the $SU(p)_k$ WZW conformal field theory. Specifically,
we demonstrate the following two results:
\begin{itemize}
\item The left-moving $\widehat{\mathfrak{su}(p)}$ affine Lie algebra at level $k$ can be
constructed from the quantum mechanical operators $Z$ and $\varphi_i$.
\item The partition function of the matrix model can be computed exactly, for
all $N$, as a function of both temperature and chemical potentials for the
$SU(p)$ global symmetry. The result \eqn{ans} is an expansion in {\it Schur
polynomials} and {\it Kostka polynomials} (both of which will be defined later in the paper). 
In the large $N$ limit, the partition function is proportional to the partition
function of the chiral $SU(p)_k$ WZW model.
\end{itemize}
This second property requires some elaboration as the matrix model partition
function depends in a rather delicate way on how we take the large $N$ limit.
To recover the chiral WZW partition function --- 
also known as the vacuum character
--- one should set $N$ divisible by $p$ and subsequently take the large $N$
limit.

\para
One can also ask what happens if we take the large $N$ limit when $N=M$ mod
$p$. In this case, we show that the quantum mechanics partition function is
equal to the character of the WZW model associated to a primary in a
representation which is perhaps best described as the ``$k^{\rm th}$-fold
symmetrisation of the $M^{\rm th}$ antisymmetric representation of $SU(p)$". In
terms of Young diagrams, this representation is
\be   M\left\{\begin{array}{c} \\ \\ \end{array}\right. \!\!\!\overbrace{
\raisebox{-3.1ex}{\yng(5,5,5)}}^k\label{rep}\ee
%
%
%
%
\para

\subsubsection*{Relationship to Chern-Simons Theory}

The connection between the matrix model \eqn{matrix} and the WZW model is not
coincidental; both are related to Chern-Simons theories. Before we derive the
results above, we will first explain why the results described in this paper
are not unexpected.

\para
The  matrix model \eqn{matrix} describes the dynamics of vortices in a $d=2+1$
dimensional Chern-Simons theory, coupled to non-relativistic matter.  This
Chern-Simons theory has gauge group and levels
\be U(p)_{k',k}= \frac{U(1)_{k'} \times SU(p)_k}{{\bf Z}_p}\label{thisnow}\ee
where $k'=(k+p)p$.

\para
The relationship between the matrix model and non-Abelian vortex dynamics was
explained in \cite{us2}, following earlier work on vortices in the $p=1$
Abelian Chern-Simons theory \cite{unknown,us1}.
 The vortices sit in a harmonic trap which forces them to cluster around the
origin, where they form a droplet of size $\sim \sqrt{N}$.  Outside this
region, the gauge group $U(p)$ is broken; inside it is unbroken.

\para
The upshot is that the solitonic vortex provides a way to engineer Chern-Simons
theory on a manifold with boundary, where the role of the boundary is played by
the edge of the vortex. It is well known that the gapless excitations of the
Chern-Simons theory are chiral edge modes, described by a WZW model with
algebra $U(p)_{k',k}$ \cite{wittenknot,israeli}. The advantage of the present
set-up is that we can identify the microscopic origin of these edge modes as
the excitations of the vortices.  These excitations are captured by the matrix
model \eqn{matrix}.

\para
The vortex perspective also provides a way to understand the delicate manner in
which we should take the large $N$ limit. It was shown in \cite{us2} that the
vortices have a unique, $SU(p)$ singlet, ground state only when $N$ is
divisible by $p$. As we described above, with this restriction in place, the
large $N$ limit of the partition function coincides with the partition function
of the WZW model.

\para
In contrast, when $N =M$ mod $p$, the ground state of the vortices is not
unique; rather, it transforms in the representation \eqn{rep}. This explains
why taking the large $N$ limit keeping $N =M$ mod $p$ results in the character
of the Kac-Moody algebra associated to this representation.

\subsubsection*{Relationship to the Quantum Hall Effect}

Our original interest in the matrix model \eqn{matrix} was through its
connection to the quantum Hall effect. When $p=1$, it  reduces to the matrix
model introduced by Polychronakos to describe Laughlin states at filling
fraction $\nu = 1/(k+1)$ \cite{alexios}. (This matrix model was inspired by an earlier approach 
using non-commutative geometry \cite{susskind}.) With $p\geq 2$, the matrix model
describes a class of non-Abelian quantum Hall states with filling fraction $\nu
= p/(k+p)$ \cite{us2}.  These lie in the same universality class as states
previously introduced by Blok and Wen \cite{bw}.

\para
There is a deep connection between the bulk properties of quantum Hall states
and the $d=1+1$ conformal field theory which describes the dynamics of the edge
modes. This connection was first highlighted in \cite{mr} where it was shown
that the bulk wavefunction can be reconstructed as a CFT correlation function.
This  relationship was subsequently  used to derive several interesting
non-Abelian quantum Hall states \cite{mr,rr,ard}.

\para
However, one can also go the other way. Starting from a quantum Hall
wavefunction, one can enumerate its full set of excitations. These can then be
matched to the excitations of the boundary conformal field theory. This was
first done by Wen for Abelian quantum Hall states \cite{wenedge,wen2} and later
extended to a  number of paired, non-Abelian quantum Hall states in \cite{mil}.

\para
The connection between the matrix model \eqn{matrix}  and the WZW model
highlighted in this paper falls naturally into this larger quantum Hall
narrative. Indeed, it is known that the Blok-Wen states --- which are the
ground states of our matrix model --- can be reconstructed from correlation
functions in the $U(p)_{k',k}$ WZW model \cite{bw,us2}. The results of this paper
can be thought of as a derivation of the converse story: the excitations of the
matrix model coincide with those of the boundary CFT.

\para
The excitations arising from the $p=1$ matrix model were previously shown to
coincide with those of a chiral boson \cite{edge}.
 For $p\geq 2$, the story is much richer as the  partition function now depends
on both temperature and chemical potentials for the $SU(p)$ flavour symmetry.
Nonetheless, our results show that the excitations above the quantum Hall state
do coincide with those of the boundary conformal field theory.

\subsubsection*{The Plan of the Paper}

This paper contains two main results. In Section \ref{currentsec} we construct
the Kac-Moody current algebra from the quantum mechanics. In Section
\ref{partsec} we compute the partition function of the matrix model and explain
how to take the large $N$ limit.

\para
The computation of the partition function involves a number of results from the
theory of symmetric functions. In an attempt to make this paper self-contained,
we have included in Section \ref{symsec} a review of the properties  of  Schur,
Hall-Littlewood and Kostka polynomials, which are the lead  characters in our
story. Appendix \ref{kostkasec} contains further details about Kostka
polynomials. Other appendices describe our conventions for affine Lie algebras
and the details of the current algebra computation.

\section{The Current Algebra}\label{currentsec}

The purpose of this paper is to explain how the  $N\rightarrow \infty$ limit of the matrix model \eqn{matrix} is related to the $d=1+1$ WZW conformal field theory. The smoking gun for the emergence of a WZW model is, of course, a current algebra. In this section we will show how to construct such an algebra from the matrix model degrees of freedom $Z$ and $\varphi_i$. 

\para
The key point is that the  $U(N)$  gauge symmetry ensures that $Z$ and $\varphi$  are not independent. In particular, Gauss' law of the  matrix model \eqn{matrix} constrains the degrees of freedom to obey
\be [Z,Z^\dagger] + \sum_{i=1}^p \varphi_i\varphi_i^\dagger = (k+p){\bf 1}_N\label{constraint}\ee
We'll see that the current algebra arises, in part, due to these constraints. 

\para 
Both the classical and quantum matrix models exhibit the Kac-Moody algebra.  The difference between the two appears  only to be a shift of the level. We will prove that the classical matrix model has an $\widehat{\mathfrak{su}(p)}$ algebra at level $k+p$. In the quantum theory we find level $k$. However, the extra complications in the quantum theory mean that we have been unable to complete the proof of the existence of the algebra; we rely on two conjectured identities which we present below.

\para 
This shift of the level can already be seen in the quantum version of the constraint equation \eqn{constraint}. In the quantum theory, the individual matrix and vector entries $Z_{ab}$ and $\varphi_{i\,a}$ become operators, obeying the canonical commutation relations
\be [Z_{ab},Z^\dagger_{cd} ] = \delta_{ad}\delta_{bc}\ \ \ {\rm and}\ \ \ [\varphi_{i\,a},\varphi_{j\,b}^\dagger]= \delta_{ab}\delta_{ij}\label{commute}\ee
We choose a reference state $|0\rangle$ obeying $Z_{ab}|0\rangle = \varphi_{i\,a} |0\rangle=0$ and construct a Hilbert space by acting with $Z^\dagger_{ab}$ and $\varphi^\dagger_{i\,a}$. The quantum version of the Gauss' law constraint \eqn{constraint} is interpreted as the requirement that physical states are $SU(N)$ singlets; this can be written in normal ordered form as
\be :[Z, Z^\dagger]: + \sum_{i=1}^p\varphi_i\varphi_i^\dagger  = (k+p) {\bf 1}_N\label{qgauss}\ee
Here the $:\ :$ determines the order in which  operators appear -- with $Z$ moved to the right -- but not the way that $U(N)$ group indices are contracted; this is determined by the matrix commutator $[\,,\,]$. Meanwhile, the level determines the charge under $U(1)\subset U(N)$ that physical states must  carry. Taking the  trace of Gauss' law, and using the commutation relations \eqn{commute}, gives
\be \sum_{a=1}^N \sum_{i=1}^p\varphi_{i\,a} \varphi^\dagger_{i\,a} = (k+p)N \ \ \ \Rightarrow\ \ \ \sum_{a=1}^N \sum_{i=1}^p\varphi^\dagger_{i\,a} \varphi_{i\,a} = kN\label{traceconstraint}\ee
We will see below that a similar normal ordering issue shifts the level of the Kac-Moody algebra.

\subsection{The Currents}

It is straightforward to construct generators of the {\it positive graded} current algebra in the matrix model. The problem factorises into $U(1)$ and $SU(p)$ parts. 
The  $U(1)$ currents are simply
\be \tilde{\cal J}^m = {\rm tr}\,Z^m\nn\ee
while the $SU(p)$ adjoint-valued currents are
\be
\tilde{\cal J}_{ij}^m = i\left( \varphi_i^\dagger Z^m \varphi_j - \frac{1}{p} \delta_{ij}\, \varphi_k^\dagger Z^m \varphi_k \right) \nn\ee
Here $i,j,k=1,\ldots,p$ are flavour indices, while $m\geq 0$ denotes the grading. 

\para
It is simple to show that the commutators \eqn{commute} imply that these currents give a representation of half of the Kac-Moody algebra,
\be
[\tilde{\cal J}_{ij}^m, \tilde{\cal J}_{kl}^n] &=& i\left( \delta_{il} \tilde{\cal J}_{kj}^{m+n} - \delta_{kj} \tilde{\cal J}_{il}^{m+n} \right)
\label{gradedlie}\ee
while $[\tilde{\cal J}^m, \tilde{\cal J}^n]= [\tilde{\cal J}^m, {\cal J}_{ij}^n] = 0$. This holds for any $N$. This same expression holds in both the quantum theory and the classical theory where, in the latter, the commutation relations \eqn{commute} should be replaced by classical Poisson brackets.

\para
While the result \eqn{gradedlie} is heartening, our interest really lies in the full Kac-Moody algebra and, in particular, the central extension term. Here we will see the difference between classical and quantum theories.  

\para
The central charge of the  $U(1)$ current is harder to pin down due to a possible rescaling. For this reason, we 
focus on the $SU(p)$ currents. Here too there is a normalisation issue, but one that will turn out to be uniquely fixed. To this end, we rescale the positive-graded currents
\be {\cal J}^m_{ij} = \left(\frac{(k+p)N}{p}\right)^{-m/2}  \tilde{\cal J}_{ij}^m\ \ \ \ \ \ \ \  m \geq 0\nn\ee
Note that these still obey the algebra \eqn{gradedlie} since the overall scaling is a power of $m$. We will see that only these rescaled currents will give rise to the full Kac-Moody algebra. We then define the negative graded currents as
\be  {\cal J}_{ij}^m = {\cal J}^{|m|\,\dagger}_{ji} \ \ \ \ \ \ \ \ \ \ \ \ \ \ m<0\nn\ee
and similarly for $\tilde{\Icurr}$.  These too obey the graded Lie algebra \eqn{gradedlie} if we restrict to $m,n<0$.

\para
Of course, the central term only arises when we consider mixed commutators  of the form $[{\cal J}_{ij}^m, {\cal J}_{kl}^{-n}]$ with $m,n>0$. These are trickier to compute because now the constraint \eqn{constraint} comes into play. However, things simplify somewhat in the  $N \rightarrow \infty$  limit. 
We will show that the currents obey the Kac-Moody algebra
\be
[{\cal J}_{ij}^m, {\cal J}_{kl}^n] &\sim& i( \delta_{il} {\cal J}_{kj}^{m+n} - \delta_{kj} {\cal J}_{il}^{m+n} ) + k m \ \delta_{m+n,0} \ \left(\delta_{jk} \delta_{il} - \frac 1 p \delta_{ij}\delta_{kl}\right)
\ \ \ \ \ \label{kmalg}\ee
%
%
Here $\sim$ means up to $1/N$ corrections. Moreover, the operators in this equation should act on states that are constructed from the vacuum $|0\rangle$  by acting with fewer than ${\cal O}(N)$ creation operators.

\para
The rest of this section is devoted to the derivation of \eqn{kmalg}. (We also show this structure arises perturbatively, in a sense made clear in the appendix, in the Poisson brackets of the {\em classical} theory. In that setting we obtain an algebra at the unshifted level $k + p$.)

\subsection{Deriving the Kac-Moody Algebra}

The novelty in deriving \eqn{kmalg} arises from the commutator $[Z,Z^\dagger]$ terms between currents travelling in opposite directions. We take $m,n>0$ and look at 
\be [\tilde{\Icurr}_{ij}^m,\tilde{\Icurr}_{kl}^{-n}] &=& [\varphi_i^\dagger Z^m \varphi_j, \varphi_k^\dagger Z^{\dagger n} \varphi_l] \nn\\ &=& \delta_{jk} \varphi_i^\dagger Z^m Z^{\dagger n} \varphi_l - \delta_{il} \varphi_k^\dagger Z^{\dagger n} Z^m \varphi_j 
 + \varphi_{ia}^\dagger \varphi_{kb}^\dagger [Z^m_{ac},Z^{\dagger n}_{bd}] \varphi_{jc} \varphi_{ld} \nn\\
&=& \delta_{jk}\varphi_i^\dagger [ Z^m , Z^{\dagger n} ] \varphi_l +    \delta_{jk} \varphi_i^\dagger Z^{\dagger n} Z^m \varphi_l \nn\\ && \ \ \ \ \  - \delta_{il} \varphi_k^\dagger Z^{\dagger n} Z^m \varphi_j  + \varphi_{ia}^\dagger \varphi_{kb}^\dagger [Z^m_{ac},Z^{\dagger n}_{bd}] \varphi_{jc} \varphi_{ld}
\label{asleeptoparis} \ee
All $U(N)$ group indices are contracted in the obvious manner apart from in the final term where we've written them explicitly.

 \para
 Our first goal is to simplify the two commutators in this expression. We deal with them in turn.
 For the first, we write
\be [Z^m,Z^{\dagger n}] = \sum_{r=0}^{m-1}\sum_{s=0}^{n-1} Z^rZ^{\dagger s}[Z,Z^\dagger] Z^{\dagger n-1-s}Z^{m-1-r}\label{zzzz}\ee
We're going to replace the factor $[Z,Z^\dagger]$ appearing here with some combination of $\varphi_i\varphi_i^\dagger$ using Gauss' law \eqn{qgauss}. However, Gauss' law is not an identity between operators; instead it holds only when evaluated on physical states $|{\rm phys}\rangle$. If we don't include the normal ordering in \eqn{qgauss}, then the constraint is written as 
\be \left([Z,Z^\dagger] + \varphi_i\varphi_i^\dagger\right)|{\rm phys}\rangle = (k+p+N)|{\rm phys}\rangle\label{qqgauss}\ee
To this end, we  consider the operator $\varphi_i^\dagger [ Z^m , Z^{\dagger n} ] \varphi_l$ acting on a physical state. Then, after some manipulation, we can use \eqn{qqgauss} to write
\be \varphi_i^\dagger [ Z^m , Z^{\dagger n} ] \varphi_l |{\rm phys}\rangle &=& \displaystyle \sum_{r=0}^{m-1}\sum_{s=0}^{n-1}\varphi_i^\dagger  Z^{r} Z^{\dagger s} [Z,Z^\dagger] Z^{\dagger n-1-s} Z^{m-1-r}  \varphi_l  |{\rm phys}\rangle \nn\\
&=&\displaystyle \sum_{r=0}^{m-1}\sum_{s=0}^{n-1} \varphi_i^\dagger Z^{r} Z^{\dagger s} (k +p - \varphi_{i'}\varphi_{i'}^\dagger) Z^{\dagger n-1-s} Z^{m-1-r}  \varphi_l   |{\rm phys}\rangle \nn\\
&=& -  \displaystyle \sum_{r=0}^{m-1}\sum_{s=0}^{n-1} \varphi_i^\dagger Z^{r} Z^{\dagger s} \varphi_{i'}\varphi_{i'}^\dagger Z^{\dagger n-1-s} Z^{m-1-r}  \varphi_l  |{\rm phys}\rangle  \label{traintoparis}\\
&& \qquad \qquad\ \ \ \ + (k+p)n \displaystyle \sum_{r=0}^{m-1} \varphi_i^\dagger Z^{r} Z^{\dagger n-1} Z^{m-1-r}  \varphi_l  |{\rm phys}\rangle \nn\ee
This term above proportional to $(k+p)n$ is key:  it will become the central term in the algebra. We'll come back to this shortly. Meanwhile, the first term combines nicely with the second commutator in \eqn{asleeptoparis}. Using the expansion \eqn{zzzz}, it can be written as
\be
\varphi_{ia}^\dagger \varphi_{kb}^\dagger [Z^m_{ac},Z^{\dagger n}_{bd}] \varphi_{jc} \varphi_{ld} &=& \varphi_{ia}^\dagger \varphi_{kb}^\dagger \left( \displaystyle \sum_{r=0}^{m-1}\sum_{s=0}^{n-1} (Z^{r} Z^{\dagger s})_{ad} (Z^{\dagger n-1-s} Z^{m-1-r})_{bc} \right) \varphi_{jc} \varphi_{ld} \nn \\
&=& \displaystyle \sum_{r=0}^{m-1}\sum_{s=0}^{n-1} ( \varphi_{i}^\dagger Z^{r} Z^{\dagger s} \varphi_{l}) ( \varphi_{k}^\dagger Z^{\dagger n-1-s} Z^{m-1-r}  \varphi_{j}) + \delta_{kl} \Big( \cdots \Big)
\nn\ee
where the term proportional to $\delta_{kl}$ arises from commuting $\varphi^\dagger_{kb}$ past $\varphi_{ld}$. It can be neglected simply because  we are ultimately  interested in the $kl$-traceless part of this expression. 

\para
The four-$\varphi$ term above is very close to that appearing in \eqn{traintoparis}; it differs only in its $U(p)$ indices and overall sign. Adding the two together gives
\be  \Theta &=& \sum_{r=0}^{m-1}\sum_{s=0}^{n-1} \left[ ( \varphi_{i}^\dagger Z^{r} Z^{\dagger s} \varphi_{l})  ( \varphi_{k}^\dagger Z^{\dagger n-1-s} Z^{m-1-r}  \varphi_{j}) \right. \nn\\ &&\ \ \ \ \ \ \ \ \ \ \ \ \ \ \ 
- \delta_{jk}  \left.(\varphi_i^\dagger Z^{r} Z^{\dagger s} \varphi_{i'}) (\varphi_{i'}^\dagger Z^{\dagger n-1-s} Z^{m-1-r}  \varphi_l)  \right]\nn \ee
We can manipulate the index structure to exploit this similarity: we separate the two double sums into their trace and traceless parts with respect to $\delta_{il}$, $\delta_{jk}$, $\delta_{ii'}$ and $\delta_{i'l}$. Doing this we find that the products of traces cancel between the two pairs, as do half of the trace-traceless terms, leaving only  traceless-traceless terms which we neglect on the grounds that they are subleading in the large $N$ limit. We're left with 
\be \Theta &\sim&   \frac{1}{p} \sum_{r=0}^{m-1}\sum_{s=0}^{n-1} \left[\delta_{il} ( \varphi_{i'}^\dagger Z^{r} Z^{\dagger s} \varphi_{i'}) ( \varphi_{k}^\dagger Z^{\dagger n-1-s} Z^{m-1-r}  \varphi_{j}) \right. \nn\\ &&\ \ \ \ \ \ \ \ \ \ \ \ \ \ \  -  \left.\delta_{jk}( \varphi_{i'}^\dagger Z^{r} Z^{\dagger s} \varphi_{i'}) ( \varphi_{i}^\dagger Z^{\dagger n-1-s} Z^{m-1-r}  \varphi_{l})\right] \nn\ee
%
%
%
%
The first term above and the  third term in \eqn{asleeptoparis} are both proportional to $\delta_{il}$; similarly, the second term above and  the second term in \eqn{asleeptoparis} are both proportional to $\delta_{jk}$.  In each case, the two terms combine together in the large $N$ limit. This follows from the following identity:

\para
{\bf Identity 1:} For $m\ge n$,
\be  
\varphi_i^\dagger Z^{\dagger n} Z^m \varphi_l -  \frac 1 p \displaystyle \sum_{r=0}^{m-1}\sum_{s=0}^{n-1}( \varphi_{i'}^\dagger Z^{r} Z^{\dagger s} \varphi_{i'})(\varphi_{i}^\dagger Z^{\dagger n-1-s} Z^{m-1-r}  \varphi_l)  \sim  \left( \frac{(k+p)N}{p} \right)^n \varphi^\dagger_i Z^{m-n} \varphi_l  \nn\ee
where $\sim$ again means up to $1/N$ corrections. Further, we are neglecting a trace, proportional to $\delta_{il}$ on both sides. 
 A similar expression holds when $n>m$. 
 
\para
The proof of this identity in the classical theory is already somewhat involved, so we relegate it to Appendix \ref{idsec}. We have not been able to extend the proof to the quantum case where the additional commutators \eqn{commute} make it much more challenging, though we have checked it for small $n$ and $m$. In what follows, we will make the natural assumption that this identity generalises directly to the quantum case. 
 
\para
It remains only to discuss the second term in \eqn{traintoparis}; this is our central term. We again decompose it into the trace and traceless components with respect to the $i,l$ indices. At large $N$, the traceless component is subleading; we have
\be
(k+p) n \delta_{jk} \displaystyle \sum_{r=0}^{m-1} \varphi_i^\dagger Z^{r} Z^{\dagger n-1} Z^{r-1-r}  \varphi_l \sim \frac{(k+p)n}{p} \delta_{jk} \delta_{il} \displaystyle \sum_{r=0}^{m-1} \varphi_{i'}^\dagger Z^{r} Z^{\dagger n-1} Z^{m-1-r}  \varphi_{i'}
\nn\ee
To proceed, we need a second large $N$ identity. This time the identity takes a different form in the classical and quantum theories. Evaluated on classical matrices, the identity reads

\para
{\bf Identity 2 (Classical Version):} For $m \ge n$,
\be
\sum_{r=0}^{m-1} \varphi_{i'}^\dagger Z^{r} Z^{\dagger n-1} Z^{m-1-r}  \varphi_{i'} \sim p \left( \frac {(k+p)N}{p} \right)^n  \delta_{mn}
\nn\ee
We present a proof of this identity in Appendix \ref{idsec}.

\para
Meanwhile, in the quantum theory there is an extra term which arises due to the shift $k+p\rightarrow k$ seen in \eqn{traceconstraint}. The corresponding large $N$ identity now reads

\para
{\bf Identity 2 (Quantum Version):} For $m \ge n$,
\be
\sum_{r=0}^{m-1} \varphi_{i'}^\dagger Z^{r} Z^{\dagger n-1} Z^{m-1-r}  \varphi_{i'} \sim p \left( \frac {(k+p)N}{p} \right)^n \left(1 - \frac{p}{k+p} \right) \delta_{mn}
\nn\ee
We do not have a general proof of this result in the quantum theory. Nonetheless, as before the existence of the new factor can be checked numerically in a number of simple examples.

\para
Putting all of this together, we arrive at our final result. In the large $N$ limit,  up to terms proportional to $\delta_{ij}$ and $\delta_{kl}$, we have
\be
[\varphi_i^\dagger Z^m \varphi_j, \varphi_k^\dagger Z^{\dagger n} \varphi_l] &\sim& \left( \frac{(k+p)N}{p} \right)^n \Bigg[ \delta_{jk} \varphi^\dagger_i Z^{m-n} \varphi_l - \delta_{il} \varphi^\dagger_k Z^{m-n} \varphi_j +  k n \ \delta_{mn} \  \delta_{jk} \delta_{il} \Bigg]
\nn\ee
Written in terms of currents, this is equivalent to the Kac-Moody algebra \eqn{kmalg}. 


\section{The Partition Function}\label{partsec}

In this section, we compute the partition function of the matrix
model.
In the limit of large particle number, $N\rightarrow\infty$, we will
show that this partition function is proportional to a  character of the chiral
$\hat{A}_{p-1}$ current algebra at level $k$ \eqn{kmalg}.
\para
There is a well-established machinery for solving matrix models in the
$N\rightarrow \infty$ limit;
the usual route is through the  path integral which, at large $N$,
can typically be evaluated by finding an appropriate saddle point for the
Wilson lines arising from the gauge field $\alpha$. Here we will do
better and compute the partition function exactly for all values of
$N$.  The resulting formula for the partition function, given in equation
(\ref{ans}), can
then be analysed directly in the large $N$ limit. However, the nature
of this limit is subtle; in particular, it depends
on the value of $N$ modulo $p$, and
does not seem to have a direct interpretation in terms of a saddle point of
the original matrix integral\footnote{If one tries to
  take the standard large $N$ approach, the hurdle is to find a correct
way to implement the level constraint \eqn{traceconstraint}. One can show that
integrating out the fundamental matter $\varphi_i$ results in a Wilson line for
the $SU(N)$ gauge field $\alpha$ which sits in the $kN^{\rm th}$  symmetric
representation \cite{polywilson}. Because this representation scales with $N$,
it  shifts the saddle point in a complicated manner. We have not been able to
evaluate the partition function in this approach.}.

\para
In fact, our formula
for the partition function of the matrix model can be related 
\cite{morekostka,ny1,ny2} to the 
partition function of a certain integrable lattice model in two
dimensions which gives rise to conformal field theory with affine Lie
algebra symmetry in the
continuum limit. This limit has been studied in detail
in \cite{morekostka}, and the results therein lead to a closed formula
for the large $N$ limit of the matrix model partition function as an
affine character.
\para
Our partition function will depend on both the (inverse) temperature $\beta$ and the chemical potentials $\mu_i$ for the $U(1)$ Cartan elements of
the $SU(p)$ global symmetry. Including these chemical potentials, the
Hamiltonian for the matrix model \eqn{matrix} is
\be H = \omega\,{\rm tr} Z^\dagger Z - \sum_{i=1}^p \mu_i
\varphi^\dagger_i\varphi_i\label{ham}\ee
The Hamiltonian is trivial: it
simply counts the number of $Z^\dagger$ and $\varphi_i^\dagger$ excitations,
weighted by $\omega$ and $\mu_i$ respectively. Evaluated on any physical state,
the Hamiltonian gives
\be H|{\rm phys}\rangle = \left( \omega \Delta - \sum_{i=1}^p \mu_i
J_i\right)|{\rm phys}\rangle\nn\ee
where the quantum numbers $\Delta$ and $J_i$ are integers labelling each state.

\para
Our interest is in the partition function
\be {\cal Z}(q,x_i) = {\rm Tr}\, e^{-\beta H}= {\rm Tr}\,q^\Delta \,
\prod_{i=1}^p x_i^{J_i}\nn\ee
where ${\rm Tr}$ is the trace over all states, $q=e^{-\beta\omega}$ and $x_i=
e^{\beta\mu_i}$.

\para
All the complexity in the problem lies not in the Hamiltonian, but instead in
the non-trivial structure of the physical Hilbert space originating in the constraints imposed by
the $U(N)$ gauge symmetry.
 Our strategy is to first enumerate all gauge non-invariant states and only
later project onto the gauge invariant subset. With this in mind, we introduce
further fugacities for each Cartan element of the gauge symmetry,
$U(1)^N\subset U(N)$. We call these fugacities $\omega_a$ with $a=1,\ldots,N$.
\para
If we ignore the restrictions of gauge invariance, then the Hilbert space is
simple to define: it consists of any number of  $Z_{ab}^\dagger$ or
$\varphi_{a\,i}^\dagger$ operators acting on $|0\rangle$.
Let's deal with each species of operator in turn.
The $Z$ operators lie in the adjoint representation of $U(N)$ and are singlets
under $SU(p)$. They carry quantum numbers of $\omega_a^{+1}\omega_b^{-1}$
(for some $a\neq b)$ and $\Delta = 1$. Taking the trace over states of the form
$Z^{\dagger\,r}|0\rangle$ for all possible $r$ gives the contribution to the
partition function of the form
\be {\cal Z}_Z = \prod_{a,b=1}^N\frac{1}{1-q\omega_a/\omega_b}\label{zz}\ee
Meanwhile, the $\varphi$ operators transform in the fundamental of both $U(N)$
and $SU(p)$. This means that they come with a factor $\omega_a^{+1}x_i^{+1}$ for
some $a$ and $i$. They have $\Delta=0$. Taking the trace over states of the
form $\varphi^{\dagger\,r}|0\rangle$ gives the contribution to the partition
function
\be {\cal Z}_\varphi = \prod_{a=1}^N\prod_{i=1}^p\frac{1}{1-\omega_a
x_i}\label{zphi}\ee
We now impose the requirements of gauge invariance. The physical states must be
$SU(N)$ singlets. Further, the level constraint \eqn{traceconstraint} requires
that physical states carry charge $k$ under the $U(1)$ centre of
$U(N)$ but are singlets under $SU(N)\subset U(N)$. This can be
imposed by contour integration, giving us the expression
\be {\cal Z}(q,x_i) = \frac{1}{N!}\left(\prod_{a=1}^N \frac{1}{2\pi i}\oint
\frac{d\omega_a}{\omega_a^{k+1}}\right)\prod_{b\neq c}
\left(1-\frac{\omega_b}{\omega_c}\right)\,{\cal Z}_Z\,{\cal
Z}_\varphi\label{z}\ee
where the contour of integration 
is the unit circle in the complex plane for each integration
variable. 
Here the product factor arises from the Haar measure on the group
manifold of $U(N)$. The factor of $\omega^{k+1}$ in the denominator 
ensures that the only contributions we
pick up in the contour integral are those with correct overall charge.

\para
Our strategy for evaluating the partition function will be to
expand the integrand
of (\ref{z}) in a suitable basis of polynomials. The
integration variables $w_{a}$ and the fugacities $x_{i}$ are
invariant under permutations corresponding to the Weyl groups of $U(N)$
and $SU(p)$ respectively. This means that the partition function can be
expanded
in terms of symmetric polynomials.
Before proceeding we pause to review some elementary facts about these
functions.

\subsection{A Digression on Symmetric Functions}\label{symsec}

In this subsection we review some standard facts about symmetric
functions. For further details and proofs of the statements reviewed
below see \cite{Mac}. As symmetric functions are labelled by
partitions we will begin by reviewing basic features of the latter.
\para
A {\em partition} $\lambda$ is a non-increasing
sequence of non-negative integers,
\be
\lambda_{1}\geq \lambda_{2}\geq\lambda_{3}\ldots\geq \lambda_{\ell(\lambda)}
> \lambda_{\ell(\lambda)+1}=0 \nn \ee
The number $\ell(\lambda)$ of non-zero elements in the sequence
is called the {\it length} of the partition. The sum of all the elements,
$|\lambda| = \sum_{i \geq 1}
\lambda_i$, is called the {\it weight} of the partition. We will write $\mathcal{P}$
for the set of all partitions.

\para
The {\it multiplicity} $m_{j}(\lambda)$ of the
positive integer $j$ is the number of times that $j$ appears in the partition
$\lambda$; i.e.
\be
m_{j}(\lambda)=\left|\{i\geq 1 : \lambda_{i}=j \}\right|
\nn \ee
We can specify a partition either by listing its non-zero parts,
$\lambda=(\lambda_{1},\lambda_{2},\ldots,\lambda_{\ell(\lambda)})$ or
by specifying multiplicities. For example the partition $(7,5,5,3,3)$
can alternatively be written as $(7^{1},5^{2},3^{2})$ where the
exponent of each entry indicates its multiplicity. We will use this
notation extensively below.

\para
A partition $\lambda$ can be represented graphically by a
{\it Young diagram} $\mathbb{Y}(\lambda)$. This is an
array of boxes where the $i^{\rm th}$ row contains $\lambda_i$ boxes. Each row is
aligned so that the left-most boxes sit under each other. For example, the
Young diagram for the partition $(7^1,5^2,3^2)$ looks like this:
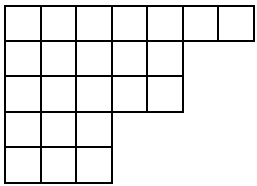
\be \yng(7,5,5,3,3)\nn\ee
Concretely the set $\mathbb{Y}(\lambda)$ contains boxes $x=(r,s)$
labelled by their coordinate $r$ and $s$ specifying the row and column
respectively of the diagram relative to the top left hand corner of
the diagram. The Young diagram $\mathbb{Y}(\lambda)$ therefore
contains boxes $x=(r,s)$ with $r=1,\ldots, \ell(\lambda)$ and, for
each value of $r$, $s=1,\ldots,\lambda_{r}$.
\para
The {\em transpose} $\lambda^{T}$
of the partition $\lambda$ is obtained by interchanging the rows and
columns of the Young diagram $\mathbb{Y}(\lambda)$. Explicitly the
non-zero parts of $\lambda^{T}$ are
\be
\lambda^{T}_{i} = \left|\{j\geq 1: \lambda_{j}\geq i\}\right|
\nn
\ee
for $i=1,\ldots,\ell(\lambda^{T})=\lambda_{1}$. For example, $(7^1,5^2,3^2)^T =
(5^3,3^2,1^2)$. Finally, we  also define the function $n: \mathcal{P}\rightarrow
\mathbb{Z}_{\geq 0}$ by
\be n[\lambda] = \sum_{i\geq 1} (i-1)\lambda_{i} \nn \ee

\para
We now turn to symmetric functions.
Let $X=\{x_{1},\ldots,x_{n}\}$ denote a set of $n$ variables. A
symmetric function $f(X)=f(x_{1},x_{2},\ldots,x_{n})$ is any
polynomial of the $x_{i}$ invariant under the action of the
permutation group $S_{n}$ acting on the variables $X$, so 
\be
f\left(x_{\sigma(1)},x_{\sigma(2)},\ldots,x_{\sigma(n)}\right)=
f(x_{1},x_{2},\ldots,x_{n})\qquad{} \qquad{} \forall \sigma\in S_{n}
\nn \ee
We will frequently use the shorthand notation
$\sigma \{g(X)\}=g(x_{\sigma(1)},x_{\sigma(2)},\ldots,x_{\sigma(n)})$ for
the action of a permutation $\sigma\in S_{n}$ on an arbitrary
function $g(X)$.
\para
The set of all symmetric polynomials forms a vector space. Though
infinite dimensional, it is naturally written as a direct sum of
vector subspaces of finite dimension
corresponding to symmetric polynomials of fixed degree. The basis
vectors are labelled by partitions $\mu\in \mathcal{P}$ with
at most $n$ parts: $\ell(\mu)\leq n$. In particular, one possible
choice of basis vectors are monomial symmetric functions, given by
\be
m_{\mu}(X) = \sum_{\sigma\in S_{n}/S^{\mu}_{n}}\, \sigma \left\{
x_{1}^{\mu_{1}} x_{2}^{\mu_{2}} \ldots   x_{n}^{\mu_{n}}\right\}
\label{mon} \ee
Here $S^{\mu}_{n}$ denotes the stabiliser of the monomial
$X^{\mu}=x_{1}^{\mu_{1}} x_{2}^{\mu_{2}} \ldots   x_{n}^{\mu_{n}}$ in
$S_{n}$ and thus the sum is taken over distinct permutations
$\sigma\{X^{\mu}\}=x_{\sigma(1)}^{\mu_{1}} x_{\sigma(2)}^{\mu_{2}} \ldots
x_{\sigma(n)}^{\mu_{n}}$ of the monomial $X^{\mu}$. The degree of the
monomial symmetric function $m_{\mu}(X)$ corresponds to the weight
$|\mu|$ of the permutation $\mu$. One can easily define an inner product on the
space of symmetric functions with respect to which the monomial
symmetric functions form an orthonormal basis:
\begin{eqnarray}
\langle m_{\lambda}, m_{\mu} \rangle  & \equiv &
\frac{1}{n !}\, \left(\prod_{i=1}^{n}\,\frac{1}{2\pi i}\oint_{C} \,
  \frac{dx_{i}}{x_{i}} \right)\, \, m_{\lambda}\left(X\right) \,
m_{\mu}\left(X^{-1}\right) = \delta_{\lambda,\mu}
\nn \end{eqnarray}
Here the contour of integration is the unit circle C in the complex
$x_{i}$-plane for $i=1,2,\ldots,n$ and $X^{-1}$ denotes the $n$
variables $\{x_{1}^{-1},x_{2}^{-1},\ldots, x_{n}^{-1}\}$.
\para
Another possible set of basis vectors for the space of symmetric
functions of $n$ variables is provided by the {\it Schur functions}. For each
partition $\mu\in \mathcal{P}$, we define the Schur function
\begin{eqnarray}
s_{\mu}(X) & = &  \sum_{\sigma\in S_{n}}\, \sigma \left\{
x_{1}^{\mu_{1}} x_{2}^{\mu_{2}} \ldots   x_{n}^{\mu_{n}}\, \prod_{i>j}
\,\frac{1}{\left(1-\frac{x_{i}}{x_{j}}\right)} \right\}
\nn
\end{eqnarray}
\para
Although not immediately apparent from this definition, the Schur
function $s_{\mu}(X)$, like the monomial symmetric function
$m_{\mu}(X)$, is a polynomial in the variables $X$ of degree
$|\mu|$. The significance of the Schur functions for our problem lies in their
close relation to the representation theory of the Lie algebra
$\mathfrak{u}(n)$.
The finite dimensional, irreducible representations of $\mathfrak{u}(n)$
are inherited from those of its complexification
$\mathfrak{gl}(n,{\bf C})$. Each such representation is labelled by a
partition $\lambda$ of length $\ell(\lambda)\leq n$. Equivalently, the
representation is labelled by the Young diagram
$\mathbb{Y}(\lambda)$.
As discussed in
more detail below, the Schur
function $s_{\lambda}(X)$, evaluated on the $n$ variables
$X=\{x_{1},\ldots,x_{n}\}$ is essentially the character of the corresponding
representation $R_{\lambda}$. This correspondence is a consequence of
the famous Schur-Weyl duality between the representation theory of
$\mathfrak{u}(n)$ and that of the permutation group.
\para
Like the monomial symmetric functions
discussed above, the Schur functions provide a
basis for the vector space of symmetric functions.
Indeed one can construct a ``matrix'' $K$ giving the explicit
linear transformation between these two bases by writing
\be
s_{\lambda}\left(X\right) = \sum_{\mu}\, K_{\lambda, \mu} m_{\mu}\left(X\right)
\label{COB}
\ee
Here $K_{\lambda, \mu}$ is zero unless $|\lambda|=|\mu|$.
The non-vanishing entries of $K_{\lambda, \mu}$ are all
positive integers, known as {\it Kostka numbers}. Thinking of $s_{\lambda}$
as the character of $R_{\lambda}$,
each monomial in the Schur polynomial corresponds to a weight of the
representation and the corresponding coefficient is simply the
multiplicity of this weight. More precisely, each monomial symmetric function
$m_{\mu}$ appearing on the RHS of  \eqn{COB}
corresponds to a family
of $\mathfrak{gl}(n,\mathbb{C})$ weights
related by the action of the Weyl group. The Kostka number
$K_{\lambda, \mu}$ is the precisely the common multiplicity of these weights in
$R_{\lambda}$.

\para
The Kostka numbers also have a second interpretation
in the representation
theory of $\mathfrak{u}(n)$ which will be important below.
Let ${\bf n}^{j}$ denote the $j^{\rm th}$ symmetric power of the
fundamental representation. Then
$K_{\lambda, \mu}$ is the multiplicity of the irreducible
representation $R_{\lambda}$ in the decomposition of the tensor
product
\begin{eqnarray}
\mathcal{T}(\mu) & = &
{\bf n}^{\mu_{1}}\otimes{\bf n}^{\mu_{2}}\otimes\ldots\otimes
{\bf n}^{\mu_{\ell(\mu)}}
\label{tensor}
\end{eqnarray}

\para
The Schur functions form a complete basis for the
symmetric functions in $n$ variables. They are orthonormal with
respect to a modified inner product $\langle \,, \rangle_{S}$ defined by
\be
\langle s_{\lambda}, s_{\mu} \rangle_{S}   \equiv
\frac{1}{n !}\, \left(\prod_{i=1}^{n}\,\frac{1}{2\pi i}\oint_{C} \,
  \frac{dx_{i}}{x_{i}} \right)\, \, \prod_{i\neq j}
\left(1-\frac{x_{i}}{x_{j}}\right) \, s_{\lambda}\left(X\right) \,
s_{\mu}\left(X^{-1}\right) = \delta_{\lambda,\mu}
\label{or}
\ee
 In the group theoretic context described above, this relation is just the
familiar orthogonality of $U(n)$ characters with respect to
integration over the group manifold with the Haar measure.
The completeness of the Schur functions as a basis
is expressed by the {\it Cauchy
  identity}. For any two sets of variables $X=\{x_{1},\ldots,x_{n}\}$
and $Y=\{y_{1},\ldots,y_{m}\}$ we have
\begin{eqnarray}
\prod_{i=1}^{n}\,\prod_{j=1}^{m} \, \frac{1}{1-x_{i}y_{j}} & = &
\sum_{\lambda} \, s_{\lambda}(X)s_{\lambda}(Y) \label{chy}
\end{eqnarray}
The sum on the right-hand side can be taken over all partitions
$\lambda$ as the product of Schur functions in the summand
will vanish identically
for $\ell(\lambda)>{\rm min}\{n,m\}$.
\para
As stated above, our goal will be to evaluate the matrix model
partition function by expanding the integrand of \eqn{z} in terms of
symmetric functions. Because of the presence in this integrand of the
Haar measure factor, together with the adjoint partition function
$\mathcal{Z}_{Z}$, it will be convenient to introduce yet another inner
product $\langle \, , \rangle_{P}$ on the space of symmetric functions
depending on an arbitrary complex parameter $q$.
For any two symmetric functions $f(X)$ and $g(X)$ we define
\begin{eqnarray}
\langle f, g \rangle_{P}  & \equiv &
\frac{1}{n !}\, \left(\prod_{i=1}^{n}\,\frac{1}{2\pi i}\oint_{C} \,
  \frac{dx_{i}}{x_{i}} \right)\, \, \frac{\prod_{i\neq j}
\left(1-\frac{x_{i}}{x_{j}}\right)}{\prod_{i\neq j}
\left(1-q\frac{x_{i}}{x_{j}}\right)}
 \, f\left(X\right) \,
g\left(X^{-1}\right)
\label{orthp} \end{eqnarray}
Note that our new inner product reduces to $\langle \,,\rangle_{S}$ in
the special case $q=0$. Can we find a new set of basis functions,
generalising the Schur functions, which are
orthogonal with respect to new measure with $q\neq 0$? In fact
the {\it Hall-Littlewood polynomials} have exactly this property.
Moreover, many of the properties of the Schur polynomials discussed
above are generalised in a nice way. For each partition $\lambda\in
\mathcal{P}$ we define
\begin{eqnarray}
P_{\lambda}(X;q) & = & \frac{1}{\mathcal{N}_{\lambda}}
\sum_{\sigma\in S_{n}}\, \sigma \left\{
x_{1}^{\lambda_{1}} x_{2}^{\lambda_{2}} \ldots   x_{n}^{\lambda_{n}}\,
\prod_{i>j}
\,\frac{\left(1-q\frac{x_{i}}{x_{j}}\right) }
{\left(1-\frac{x_{i}}{x_{j}}\right)} \right\}
\label{HLP}
\end{eqnarray}
The normalisation factor is given by
\be
\mathcal{N}_{\lambda} = \frac{\varphi_{n-\ell(\lambda)}\prod_{j \geq 1}
\varphi_{m_{j}(\lambda)} }{(1-q)^{n}}
\nn \ee
where
\be
\varphi_{m}=\prod_{j=1}^{m} \left(1-q^{j}\right)
\nn \ee
and $m_{j}(\lambda)$ denotes the multiplicity of the positive integer
$j$ in the partition $\lambda$ as defined above. As before,
$P_{\lambda}(X;q)$, is a homogeneous polynomial in the variables $X$ of degree
$|\lambda|$. It is useful to rewrite the definition \eqn{HLP} as
\begin{eqnarray}
P_{\lambda}(X;q) & = & \sum_{\sigma\in S_{n}/S^{\lambda}_{n}}\, \sigma \left\{
x_{1}^{\lambda_{1}} x_{2}^{\lambda_{2}} \ldots   x_{n}^{\lambda_{n}}\,
\prod_{\lambda_{i}<\lambda_{j}}
\,\frac{\left(1-q\frac{x_{i}}{x_{j}}\right) }
{\left(1-\frac{x_{i}}{x_{j}}\right)} \right\}
\label{HLP2}
\end{eqnarray}
where the sum is over distinct permutations of the
monomial
$X^{\lambda}=x_{1}^{\lambda_{1}} x_{2}^{\lambda_{2}} \ldots
x_{n}^{\lambda_{n}}$.

\para
As already mentioned, we have the orthogonality property
\be
\langle P_{\lambda}, P_{\mu} \rangle_{P} =  \frac{1}{\mathcal{N}_{\lambda}}\,
\delta_{\lambda, \mu}
\label{orth2} \ee
An even more striking fact is that, with the given
normalisation, each term in $P_{\lambda}(X;q)$ is itself a polynomial 
in the parameter $q$ with integer coefficients. One might instead choose to
normalise these functions to achieve orthonormality with
respect to the inner product $\langle \, , \rangle_{P}$;
however then the basis functions would no longer be polynomial in $q$.
\para
The completeness of the resulting basis is expressed in a generalisation
of the Cauchy identity. As before we consider two sets of variables
$X=\{x_{1},\ldots,x_{n}\}$
and $Y=\{y_{1},\ldots,y_{m}\}$. We now have
\begin{eqnarray}
\prod_{i=1}^{n}\,\prod_{j=1}^{m} \, \frac{1-qx_{i}y_{j}}{1-x_{i}y_{j}} & = &
\sum_{\lambda} \, b_{\lambda}(q)P_{\lambda}(X;q)P_{\lambda}(Y;q) \label{chy2}
\end{eqnarray}
where $b_{\lambda}(q)=\prod_{j\geq 1}\varphi_{m_{j}(\lambda)}(q)$.
From the definition (\ref{HLP}), the
Hall-Littlewood polynomial $P_{\lambda}(X;q)$
reduces to the Schur function $s_{\lambda}(X)$ for $q=0$. On setting
$q=1$ in the definition, we also find
$P_{\lambda}(X,1)=m_{\lambda}(X)$ where $m_{\mu}$ is the monomial
symmetric function defined in (\ref{mon}) above. Again we can find a
``matrix'' describing the change of basis from Schur to
Hall-Littlewood. Relation (\ref{COB}) is now generalised to
\be
s_{\lambda}\left(X\right) = \sum_{\mu}\, K_{\lambda, \mu}(q)
P_{\mu}\left(X; q\right)
\label{COB2}
\ee
For each choice of $\lambda$, $\mu\in \mathcal{P}$, the matrix elements
$K_{\lambda, \mu}(q)$ are polynomials in the
parameter $q$. They are known as
{\em Kostka polynomials} (see eg Chapter III.6 of \cite{Mac}) 
and they will play a central role in our
evaluation of the partition function. An explicit combinatoric formula
for the Kostka polynomials due to Kirillov and Reshetikhin 
is given in Appendix \ref{kostkasec}. Here we will
list some of their main features\footnote{The last two listed properties
  follow easily from the definition
(\ref{COB2}) and the third follows from the invertibility
of the change of basis proven in \cite{Mac}
(see Eqn (2.6) in Chapter III of this reference). The first two
properties are highly non-trivial and were first proven in \cite{LS}.}:
\begin{itemize}
\item{} They are polynomials in $q$ of degree $n[\mu]-n[\lambda]$ with
  leading coefficient equal to unity.
\item{} All non-zero coefficients are positive integers.
\item{} $K_{\lambda, \mu}(q)=0$ unless $|\lambda|=|\mu|$.
\item{} They reduce to the Kostka numbers for $q=1$: $K_{\lambda,
    \mu}(1)= K_{\lambda,
    \mu}$ for all partitions $\lambda$ and $\mu$.
\item{} $K_{\lambda, \mu}(0)=\delta_{\mu, \nu}$.
\end{itemize}
These properties ensure that the Kostka polynomials can be regarded as
a graded generalisation of the Kostka numbers. As the Kostka numbers
$K_{\lambda,\mu}$ count the number of
occurrences of the representation $R_{\lambda}$ in the tensor product
$\mathcal{T}(\mu)$ defined in \eqn{tensor}, the corresponding Kostka polynomial
$K_{\lambda,\mu}(q)$ receives a contribution $q^{\Delta}$ for some 
$\Delta\in\mathbb{Z}_{\geq 0}$, for each
such occurrence. Hence as we vary the partition $\lambda$, the Kostka
polynomial assigns an integer-valued ``energy'' $\Delta(\lambda,\mu)$ to each
irreducible component of the tensor product $\mathcal{T}({\mu})$.
It is useful to think of the representation space of $\mathcal{T}(\mu)$
as the Hilbert
space  of a spin chain with $\ell(\mu)$ sites with a $\mathfrak{u}(n)$
spin in the representation ${\bf n}^{\mu_{i}}$ at the $i^{\rm th}$ site.
Remarkably, the energy $\Delta$ precisely corresponds to one of the
Hamiltonians of the Heisenberg spin chain with these spins. In fact 
$\Delta$ is the essentially lattice momentum for a spin chain with
periodic boundary conditions. 
The Bethe ansatz solution of this system provides an
efficient combinatoric description of the corresponding Kostka
polynomials and leads directly to the explicit formulae given in
Appendix \ref{kostkasec}.

\para
To evaluate the partition function we will need one more class of
symmetric functions known as\footnote{These are also
sometimes referred to as {\em Milne
  polynomials} in the mathematical literature.}
 {\em Modified Hall-Littlewood polynomials} \cite{magic,kir1} $Q'_{\mu}(X;q)$.
For our purposes, this polynomial is
defined by the formula
\be
Q'_{\mu}(X;q) = \sum_{\lambda}\, K_{\lambda, \mu}(q)
s_{\lambda}\left(X\right)
\label{mhp} \ee
Importantly, this definition yields a non-zero answer for partitions
$\mu$ of any length\footnote{This follows because, although the RHS
of (\ref{mhp}) vanishes identically for partitions $\lambda$ of length
greater than $n$, there is no such constraint for the partition $\mu$
appearing in the Kostka polynomial $K_{\lambda, \mu}(t)$. 
Indeed we evaluate several examples of this type in the following 
using the combinatorial algorithm of Appendix B.}
Thus, unlike the other symmetric functions
defined above, $Q'_{\mu}(X;q)$ does not vanish identically\footnote{Note that in
  some references the definition of the modified Hall-Littlewood
  polynomial in $n$ variables is nevertheless 
restricted to the case $\ell(\mu)\leq n$.} for partitions with 
$\ell(\mu)>n$. 

\para
As discussed above, each Schur function $s_{\lambda}$ is the character of
a $\mathfrak{u}(n)$ representation $R_{\lambda}$.
Meanwhile, the Kostka polynomial
$K_{\lambda,\mu}(q)$ is non-zero only for irreducible representations
$R_{\lambda}$ occurring in the tensor product $\mathcal{T}(\mu)$
defined in \eqn{tensor}. Further, for each $\lambda$, $K_{\lambda, \mu}(q)$
receives a contribution $q^{\Delta}$ for each occurrence of the irrep
$R_{\lambda}$ in $\mathcal{T}({\mu})$ where $\Delta$ is the
appropriate spin
chain Hamiltonian. Putting these facts together we
learn that $Q'(X;q)$ has a natural interpretation as the
partition function of a spin chain defined on the tensor product space
$\mathcal{T}(\mu)$.
\paragraph{}
Using the properties of the Schur functions and Kostka
polynomials, we see that $Q'_{\mu}(X;q)$ is a homogeneous polynomial
in the variables $X=\{x_{1},\ldots,x_{n}\}$ of degree
$|\mu|$. Moreover the coefficients are themselves polynomials in $q$
with positive integer coefficients. The polynomial
$Q'_{\mu}$ has the following key property: it is adjoint to the ordinary
Hall-Littlewood polynomials $P_{\mu}$ with respect to the inner
product $\langle \, , \rangle_{S}$ for Schur functions \cite{kir1}. 
For any two sets of variables $X=\{x_{1},\ldots,x_{n}\}$
and $Y=\{y_{1},\ldots,y_{m}\}$ we have
\begin{eqnarray}
\prod_{i=1}^{n}\,\prod_{j=1}^{m} \, \frac{1}{1-x_{i}y_{j}} & = &
\sum_{\lambda} \, s_{\lambda}(X)s_{\lambda}(Y) \nn \\ 
& = & \,\, \sum_{\lambda, \rho} K_{\lambda,
  \rho}(q)\,P_{\rho}(Y;q)\,s_{\lambda}(X) 
\nn \\ 
& = & \sum_{\rho} \, Q'_{\rho}(X;q)P_{\rho}(Y;q)
\label{chy3}
\end{eqnarray}
using (\ref{COB2}) and (\ref{mhp}). 
The final sum on the RHS can be taken over all partitions $\rho$
but the summand will vanish unless $\ell(\rho)\leq m$.

\subsection{Back to the Partition Function}

We are now ready to compute the partition function $\mathcal{Z}$ defined in
\eqn{z}. The
partition function is symmetric in the $\mathfrak{u}(p)$ fugacities
$X=\{x_{1},x_{2},\ldots,x_{p}\}$ so we can expand it in terms of Schur
functions. As each Schur function corresponds to the
character of a finite-dimensional irreducible representation of
$U(p)$, the resulting expansion determines the multiplets of the
$SU(p)\subset U(p)$ global symmetry present in the matrix model spectrum.
The integrand of the partition function is also a
symmetric function of the $U(N)$ fugacities
$\Omega=\{\omega_{1}\ldots,\omega_{N}\}$ and we may thus expand it in
terms of a suitable set of basis functions.

\para
To proceed to the answer by the shortest path, we will use the Cauchy
identity in the form 
(\ref{chy3}) to expand the factor of the integrand corresponding to the
fundamental-valued fields as
\be {\cal Z}_\varphi = \prod_{a=1}^N\prod_{i=1}^p\,\frac{1}{1-\omega_a x_i} =
\sum_\lambda Q'_\lambda(X;q)P_\lambda(\Omega;q)\nn\ee
In contrast, the corresponding factor $\mathcal{Z}_{Z}$ for the
adjoint-valued field will be left unexpanded 
as part of the integration measure.   
For the next step, we use the definition of the Hall
polynomials in its second form (\ref{HLP2}) to write
\be
\frac{1}{\prod_{a=1}^{N} \omega_{a}^{k}} =
P_{(k^{N})}\left(\Omega^{-1};q\right)
\nn \ee
where, as above, $(k^{N})$ denotes the partition with $N$ non-zero
parts each equal to $k$.
The resulting integral over the variables $\Omega$ can then be written, using (\ref{orth2}),
as an inner product
\begin{eqnarray}
\mathcal{Z} & = & \sum_{\lambda}\, Q'_{\lambda}(X;q)\,\, \times \,\,
\frac{1}{N !}\, \left(\prod_{a=1}^{N}\,\frac{1}{2\pi i}\oint_{C} \,
  \frac{d\omega_{a}}{\omega_{a}} \right)\, \, \frac{\prod_{a\neq b}
\left(1-\frac{\omega_{a}}{\omega_{b}}\right)}{\prod_{a, b}
\left(1-q\frac{\omega_{a}}{\omega_{b}}\right)}
 \, P_{\lambda} \left(\Omega;q\right) \,
P_{(k^{N})} \left(\Omega^{-1};q\right) \nn \\
& = &  \sum_{\lambda}\, Q'_{\lambda}(X;q) \,\, \times \,\,
\frac{1}{(1-q)^{N}} \langle
P_{\lambda}, P_{(k^{N})}\rangle_{P} \\\nn
 &=& \frac{1}{\varphi_{N}(q)}\,\,Q'_{(k^{N})}(X;q) \nn
\nn
\end{eqnarray}
Thus our final result for the partition function of the matrix model
is
\begin{eqnarray}
\mathcal{Z} & = & \prod_{j=1}^{N} \frac{1}{(1-q^{j})}\,\,\sum_{\lambda}
K_{\lambda, (k^{N})}(q)\, s_{\lambda}(X)
\label{ans}
\end{eqnarray}
where the sum on the right-hand side runs over all partitions
$\lambda\in \mathcal{P}$. However, as explained above the summand
vanishes unless we have $|\lambda|=|(k^{N})|=kN$ and
$\ell(\lambda)\leq p$.

\subsubsection*{Ground State Energy}

Our result for the partition function \eqn{ans} holds for all positive integral
values of the level, $k$, the rank $p$ of the global symmetry
and the particle number $N$. In the remainder of this section we will
extract the ground state energy $E_{0}(k,p,N)$, which simply
corresponds to the leading power of $q$ appearing in the expansion of
the partition function for $|q|\ll1$, and compare it with our
expectations based on the analysis of the ground state given in \cite{us2}.

\para
We  begin with the abelian case $p=1$. In this case there is only
one partition $\lambda=(kN)$ which satisfies the conditions $|\lambda|=kN$,
and $\ell(\lambda)\leq 1$. In this special case, the formulae for
the Kostka polynomial given in Appendix \ref{kostkasec} simplify,
giving\footnote{Here we see explicitly that $K_{\lambda,\mu}(t)$ can be
  nonzero when $\ell(\mu)>\ell(\lambda)$ as mentioned above.}  
\be
K_{(kN),(k^{N})}(q) = q^{n\left[(k^{N})\right]} \nn
\ee
where, as above, $n[\mu]=\sum_{i\geq 1} (i-1)\mu_{i}$.  
Thus the
partition function for $p=1$ reads
\begin{equation}
\mathcal{Z}_{p=1} =x^{kN} q^{\frac{k}{2}N(N-1)}\prod_{j=1}^{N}
\frac{1}{(1-q^{j})}
\label{zp1}
\end{equation}
where $x=x_{1}$ acts as a fugacity for the $U(1)$ charge which is
fixed by the D-term constraint.
The ground state energy,
\begin{eqnarray}
E_{0}(k,1,N) & = &  \frac{k}{2}N(N-1)
\nn
\end{eqnarray}
agrees with the identification of the ground state given in
\cite{us2}. The remaining factor in (\ref{zp1}) is a plethystic exponential
accounting for excitations corresponding to
all possible products of the $N$ independent
single-trace operators ${\rm tr} (Z^{l})$ with $l=1,2,\ldots,N$. This partition
function for the $p=1$
matrix model was previously computed in \cite{edge}.

\para
The partition function for general $p\geq 1$ is somewhat richer; it also
depends on
the fugacities $x_i$ for the $SU(p)$ Cartan elements.
To understand the form of this partition function, we start by recalling that
the Kostka polynomial $K_{\lambda,
  (k^{N})}(q)$ specialises for $q=1$ to the Kostka number $K_{\lambda,
  (k^{N})}$. This in turn
coincides with the multiplicity of the irreducible representation
of $\mathfrak{u}(p)$ specified by the partition
$\lambda$ in the tensor product
\be
\mathcal{T}_{N} = {\bf p}^{k} \otimes {\bf p}^{k}\otimes \ldots
\otimes {\bf p}^{k} \nn \ee
of $N$ copies of the $k^{\rm th}$ symmetric power of the fundamental
representation ${\bf p}$. The
corresponding Kostka polynomial is a gradation of the Kostka number
where each power of $q$ appears with a non-negative integer
coefficient. As the Kostka
polynomial $K_{\lambda, (k^{N})}(q)$ appears in (\ref{ans})
multiplied by the corresponding Schur function $s_{\lambda}(X)$, we
deduce that the full spectrum of the matrix model transforms in the
reducible $\mathfrak{u}(p)$ representation $\mathcal{T}_{N}$. More
precisely, the overall prefactor of $\prod_{j=1}^{N}(1-q^{j})^{-1}$ means
we actually have an infinite number of copies of this representation.
The additional information contained in the partition function is the
energy of each irreducible component in the tensor product.
\para
To go further we will
need to use the combinatoric description of the Kostka polynomials
given in Appendix \ref{kostkasec}. We will start with the easiest case $k=1$
where an
explicit formula is available. Here we have
\begin{eqnarray}
 K_{\lambda,(1^{N})}(q) & = & q^{n\left[\lambda^{T} \right]} \,
\frac{\prod_{j=1}^{N} {(1-q^{j})}}{H(q)}
\nonumber
\end{eqnarray}
where $H(q)$ is the hook-length polynomial given by
\begin{eqnarray}
H(q) & = & \prod_{x\in \mathbb{Y}(\lambda)}\, \left(1- q^{h(x)}\right)
\nonumber
\end{eqnarray}
Here the product is over the boxes $x=(r,s)$ of the Young diagram
$\mathbb{Y}(\lambda)$ corresponding to the partition $\lambda$ and
$h(x)=\lambda_{r}+\lambda^{T}_{s}-r-s+1>0$ is the length of the hook
passing through box $x$.

\para
As we described above, $\lambda^{T}$ denotes the {\em transpose}
of the partition $\lambda$, obtained by interchanging the rows and
columns of the Young diagram $\mathbb{Y}(\lambda)$. Explicitly the
non-zero parts of $\lambda^{T}$ are
\be
\lambda^{T}_{i} = \left|\{j\geq 1: \lambda_{j}\geq i\}\right|
\nn
\ee
for $i=1,\ldots,\ell(\lambda^{T})=\lambda_{1}$. To find the
ground state energy of the model we must therefore minimise the
quantity
\be
n\left[\lambda^{T}\right] = \sum_{i\geq 1} (i-1)\lambda^{T}_{i}
\nn \ee
as we vary $\lambda$ over partitions with $|\lambda|=N$ and
$\ell(\lambda)\leq p$. These restrictions correspond to demanding that
$|\lambda^{T}|=|\lambda|=N$ and that $\lambda^{T}_{i}\leq p$ for all
$i\geq 1$. Writing $N=Lp+M$ for non-negative integers $L$ and $M<p$,
the minimum occurs for the partition $\lambda_{0}^{T}=(p^{L},M)$
corresponding to
\be
\lambda_{0}=\left((L+1)^{M}, L^{p-M}\right)
\nn
\ee
Thus the leading term in the partition sum for $|q|\ll1$ is
\be
\mathcal{Z}\simeq q^{E_{0}(N)}
 \,s_{((L+1)^{M},L^{p-M})}\left(x_{1},\ldots, x_{p}\right)
\nn
\ee
with vacuum energy
\begin{eqnarray}
E_{0}(1,p,N) & = & n\left[ (p^{L},M) \right] = \frac{1}{2}L(L-1)p + LM
\nn \end{eqnarray}
for $N=Lp+M$.
The Schur polynomial corresponds to the representation of $U(p)\simeq
U(1)\times SU(p)$ with $U(1)$ charge $N$, which coincides with the $M^{\rm th}$
antisymmetric power of the fundamental. Again, this yields complete agreement
with the
properties of the ground state discussed in \cite{us2}.
\paragraph{}
Although the formulae for the Kostka polynomials are more complicated,
the generalisation of this analysis to $k>1$ is straightforward. As we
discuss in Appendix \ref{kostkasec}, the minimum energy is obtained for the
partition
\be
\lambda_{0}=\left((kL+k)^{M}, (kL)^{p-M}\right)
\nn
\ee
and takes the value
\begin{eqnarray}
E_{0}(k,p,N) & = &  \frac{k}{2}L(L-1)p + kLM
\nn \end{eqnarray}
for $N=Lp+M$.
The resulting ground state has $U(1)$ charge $kN$ and transforms in an
irreducible representation of $SU(p)$ corresponding to a $k$-fold
symmetrisation of the $M^{\rm th}$ antisymmetric power of the fundamental
representation. This is the representation \eqn{rep} that we mentioned in the
introduction; it is
in  agreement with the results of \cite{us2}.

\subsection{The Continuum Limit}

In this section we will investigate the $N\rightarrow \infty$ limit of
the partition function \eqn{ans}. As we saw in the previous section, the
ground state energy and its quantum numbers under the global $U(p)$
symmetry depend sensitively on the value of $N$ mod $p$. This means that in
order to get
a sensible limit, we must hold this value fixed as $N\rightarrow\infty$.
Setting
\be N=Lp+M\nn\ee
 for non-negative integers $M<p$ and $L$,
we therefore take the limit $L\rightarrow \infty$ with $M$ and $p$
held fixed.

\para
It is also convenient to factorise the partition function as
\begin{eqnarray}
\mathcal{Z} & = & q^{E_{0}}w^{kN/p}\hat{\mathcal{Z}}
\nn
\end{eqnarray}
where
\begin{eqnarray}
E_{0} & = & E_{0}(k,p,N)=\frac{k}{2}L(L-1)p+kLM
\nn
\end{eqnarray}
is the ground state energy and $w=x_{1}x_{2}\ldots x_{p}$ is the
fugacity for the $\mathfrak{u}(1)$ centre of $\mathfrak{u}(p)$.
The reduced partition function
\begin{eqnarray}
\hat{\mathcal{Z}} & = &  \prod_{j=1}^{N} \frac{1}{(1-q^{j})}\,\,
q^{-E_{0}}\sum_{\lambda}
K_{\lambda, (k^{N})}(q)\, w^{-kN/p} s_{\lambda}(X)
\label{zhat1}
\end{eqnarray}
thus encodes the
energies and $\mathfrak{u}(1)$ charges of the states in the spectrum
relative to those of the ground state. As we will see,
it is $\hat{\mathcal{Z}}$ rather than the original partition function
which has a non-singular $N\rightarrow\infty$ limit.  The main result of this
section is \eqn{final} which says (roughly) that
\begin{eqnarray}
\lim_{N\rightarrow \infty} \hat{\mathcal{Z}} =
\prod_{j=1}^{\infty} \frac{1}{(1-q^{j})}\,\,\,\, \chi_{R_{k,M}}
\left(q;X\right)
\label{lovely}
\end{eqnarray}
Here the ``roughly" refers to a slight notational subtlety regarding the
difference between the $U(p)$ fugacities labelled by $X$ and the
$SU(p)$ fugacities; this will be explained below.  The key part of the result
is that $\chi_{R_{k,M}}$ denotes the character of the $\hat{A}_p$
affine Lie algebra at level $k$ 
associated to the representation $R_{k,M}$ of $SU(p)$. When $M=0$,
so $N$ is divisible by $p$, this is the vacuum character which coincides with
the partition function of the WZW model.  Meanwhile, for $M\neq 0$,
$R_{k,M}$ is the
$k$-fold symmetrisation of the $M^{\rm th}$ antisymmetric representation,
namely \eqn{rep}. The remainder of this section is devoted to the derivation of
\eqn{lovely}.
\para
As discussed above, each Schur function $s_{\lambda}(X)$ appearing in
the sum on the RHS of (\ref{zhat1}) is
the character of the irreducible
representation of $\mathfrak{u}(p)$ corresponding to the partition
$\lambda$. In the following it will be convenient to decompose the global
symmetry as $\mathfrak{u}(p)\simeq \mathfrak{u}(1)\oplus
\mathfrak{su}(p)$. Recall that the finite-dimensional, irreducible
representations of $\mathfrak{su}(p)$ are in one-to-one correspondence
with Young diagrams having at most $p-1$ rows or, equivalently with
partitions $\tilde{\lambda}$ having $\ell(\tilde{\lambda})<p$. In
contrast, representations of $\mathfrak{u}(p)$ correspond to
diagrams with at most $p$ rows or to partitions $\lambda$ with
$\ell(\lambda)\leq p$. Given an irreducible representation of
$\mathfrak{u}(p)$, we obtain a unique irreducible representation of
$\mathfrak{su}(p)$ by removing all columns of height $p$ from the
corresponding Young diagram. Similarly, for any partition $\lambda$
with $\ell(\lambda)\leq p$ we may find a unique partition
$\tilde{\lambda}$ with $\ell(\tilde{\lambda})<p$ such that
$\lambda_{i}=\tilde{\lambda}_{i}+Q$ with $i=1,\ldots,p$ for some
non-negative integer $Q$. In the following we will
abbreviate this relation as
\be
\lambda = \tilde{\lambda} + (Q^{p})
\label{lam}
\ee
The Kostka polynomial $K_{\lambda,(k^{N})}(q)$ is only non-zero if
$|\lambda|=|(k^{N})|=kLp+kM$.
Given any partition $\tilde{\lambda}$
with $\ell(\tilde{\lambda})<p$ and $|\tilde{\lambda}|\leq kN$, we may find a unique partition
$\lambda=\tilde{\lambda}+(Q^{p})$ obeying this constraint
if and only if $|\tilde{\lambda}|-kM$ is
divisible by $p$, in which case we set

\be
Q=Q(\tilde{\lambda})= kL-\frac{1}{p}\left(|\tilde{\lambda}|-kM\right)
\label{qlam}
\ee
\para
As the Schur function $s_{\lambda}(X)$ is a
homogeneous polynomial of degree $|\lambda|$ in the variables
$X=\{x_{1},x_{2}, \ldots, x_{p}\}$, we may write
\be
s_{\lambda}(X)=w^{{|\lambda|}/{p}} s_{\tilde{\lambda}}(\tilde{X})
\nn
\ee
where
\be
\tilde{X}=X/w^{{1}/{p}}=\{x_{1}w^{-{1}/{p}}, \ldots,
x_{p}w^{-{1}/{p}}\} \label{xtilde}  \ee
In particular note that, by construction,
$\tilde{x}_{1}\tilde{x}_{2}\ldots\tilde{x}_{p}=1$, which implies
$s_{\lambda}(\tilde{X})=s_{\tilde{\lambda}}(\tilde{X})$.
Using the above results, we can trade the sum over all partitions
$\lambda$ appearing in (\ref{zhat1}) for a sum over $\tilde{\lambda}$
of length $\ell(\tilde{\lambda})<p$. This gives
\begin{eqnarray}
\hat{\mathcal{Z}} & = &  \prod_{j=1}^{N} \frac{1}{(1-q^{j})}\,\,
q^{-E_{0}}\sum_{\tilde{\lambda},\,\,\, \ell(\tilde{\lambda})<p}
K_{\tilde{\lambda}+(Q(\tilde{\lambda})^{p}), (k^{N})}(q)\,
s_{\tilde{\lambda}}(\tilde{X})
\label{zhat2}
\end{eqnarray}
where $Q(\tilde{\lambda})$ is a non-negative integer
given by (\ref{qlam}) when
$|\tilde{\lambda}|=kM$ mod $p$ and $|\tilde{\lambda}|\leq kN$, and is set to zero otherwise.
\paragraph{}
It will also be useful to make the relation between Schur functions
and the characters of the simple Lie algebra $\mathfrak{su}(p)$ and
its complexification
$A_{p-1}=\mathfrak{sl}(p,\mathbb{C})$ more explicit (see Appendix \ref{liesec}
for our Lie
algebra conventions). The finite-dimensional irreducible
representations of $A_{p-1}$ are labelled by dominant integral weights
$\Lambda\in \mathcal{L}_{W}^{+}$. Each such $\Lambda$ has an expansion in
terms of the fundamental weights $\{\Lambda_{(1)},\ldots,\Lambda_{(p-1)}\}$:
\be
\Lambda=\sum_{j=1}^{p-1} \, \psi_{j}\Lambda_{(j)}
\nn
\ee
with coefficients $\psi_{j}\in \mathbb{Z}_{\geq 0}$ known as
Dynkin labels. Let $R_{\Lambda}$ denote the corresponding $A_{p-1}$
representation with representation space $\mathcal{V}_{\Lambda}$.
The
{\it character} of $R_{\Lambda}$ is a function of 
variables $Z=\{z_{1},\ldots,z_{p-1}\}$ which encodes the weights of
this representation or, equivalently, the
eigenvalues of the matrices $R_{\Lambda}(h^{i})$
representing the Cartan subalgebra generators $h^{i}$ with
$i=1,\ldots,p-1$ in the
Chevalley basis. Explicitly we define
\begin{eqnarray}
\chi_{\Lambda}(Z) & = & {\rm Tr}_{\mathcal{V}_{\Lambda}}\left[\
\prod_{j=1}^{p-1}\,
  z_{j}^{R_{\Lambda}(h^{j})}\ \right]
\nn
\end{eqnarray}
As mentioned above irreducible representations of $\mathfrak{su}(p)$
can also be labelled by partitions $\tilde{\lambda}$ with
$\ell(\tilde{\lambda})<p$. For each dominant integral weight
$\Lambda\in \mathcal{L}_{W}^{+}$,
the corresponding partition $\tilde{\lambda}(\Lambda)$ has
parts $\tilde{\lambda}_{i}=\sum_{j=i}^{p-1}\psi_{j}$. The character of
$R_{\Lambda}$ can then be related to the Schur function of the
partition $\tilde{\lambda}(\Lambda)$:
\be
\chi_{\Lambda}(Z)=s_{\tilde{\lambda}(\Lambda)}(\tilde{X})
\nn
\ee
where the variables $\tilde{X}=\{\tilde{x}_{1},\ldots,\tilde{x}_{p}\}$, obeying
$\tilde{x}_{1}\tilde{x}_{2}\ldots\tilde{x}_{p}=1$,
are related to $Z=\{z_{1},\ldots,z_{p-1}\}$ by
\begin{eqnarray}
z_{1} & = & \tilde{x}_{1} \nn \\
z_{2} & = & \tilde{x}_{1}\tilde{x}_{2} \nn \\
& \vdots & \nn \\
z_{p-1} & = & \tilde{x}_{1}\tilde{x}_{2}\ldots\tilde{x}_{p-1} \label{zx}
\end{eqnarray}
Irreducible representations $R_{\Lambda}$ of $A_{p-1}$ are further classified
by
the {\em congruence class} of the corresponding highest weight
$\Lambda$, given in terms of the Dynkin labels by the value
$\mathbb{P}(\Lambda)$ of
$\psi_{1}+2\psi_{2}+\ldots (p-1)\psi_{p-1}$ modulo $p$.
Equivalently $\mathbb{P}(\Lambda)$ is equal modulo $p$ to the
weight $|\tilde{\lambda}|$ of the partition $\tilde{\lambda}$
corresponding to $\Lambda$ or to the number of boxes in the
corresponding Young diagram. We denote by
$\mathcal{L}_{W}^{+}(\mathbb{P})$ the subset of the positive
weight lattice $\mathcal{L}_{W}^{+}$ corresponding to positive weights
in $\mathbb{P}^{\rm th}$ congruence class.
\paragraph{}
The above results mean that we can rewrite the sum over partitions
$\tilde{\lambda}$ appearing in the reduced partition function
(\ref{zhat2}) as a sum over dominant integral weights $\Lambda$ of
$A_{p-1}$ in the congruence class $\mathbb{P}(\Lambda)=kM$ mod
$p$. Our final rewriting of the reduced partition function is
\begin{eqnarray}
\hat{\mathcal{Z}} & = &  \prod_{j=1}^{N} \frac{1}{(1-q^{j})}\,\,
\sum_{\Lambda\in \mathcal{L}_{W}^{+}(kM)} K_{\Lambda}(q) \chi_{\Lambda}(Z)
\label{zhat3}
\end{eqnarray}
Here
\begin{eqnarray}
K_{\Lambda}(q) & = & q^{-E_{0}(k,p,N)}K_{\lambda(\Lambda), (k^{N})}(q)
\nonumber
\end{eqnarray}
and
\be
\lambda(\Lambda)=\tilde{\lambda}(\Lambda)+(Q(\Lambda)^{p})
\nn
\ee
where
\be
Q(\Lambda)={\rm max}\left\{ 0, kL-\frac{1}{p}\left(|\tilde{\lambda}(\Lambda)|-kM\right)\right\}
\nn
\ee
\para
We will now switch gears and consider something seemingly quite
unrelated to the above discussion; the representation theory of the
affine Lie algebra $\hat{A}_{p-1}$. (Again, see Appendix \ref{liesec} for
conventions.) We work in a Chevalley
basis with generators $\{h^{i},e^{i},f^{i}\}$ where the index $i$ now
runs from zero to $p-1$. A complete basis also includes 
the {\it derivation} or grading operator $L_{0}$ associated with the
imaginary root.
\para
The weights of any representation of $\hat{A}_{p-1}$ lie in the affine
weight lattice, whose basis vectors are the fundamental weights
$\hat{\Lambda}_{(j)}$ with $j=0,1,\ldots,p-1$.
The {\it integrable} representations of
$\hat{A}_{p-1}$ are labelled by a highest weight
\be
\hat{\Lambda} = \sum \hat{\psi}_{j}\hat{\Lambda}_{(j)}
\nn
\ee
whose $p$ Dynkin labels $\{\hat{\psi}_{j}\}$ are non-negative integers. 
Each integrable 
representation has a definite {\it level} which is a non-negative
integer given by the sum of the Dynkin indices,
\be
k=\hat{\psi}_{0}+\hat{\psi}_{1}+\ldots+\hat{\psi}_{p-1}
\nn
\ee
\para
The
resulting representations $R_{\hat{\Lambda}}$ are the affine
analogs of the finite-dimensional irreducible  representations
$R_{\Lambda}$ of the simple Lie algebra $A_{p-1}$ discussed above.
We denote the corresponding representation space
$\mathcal{V}_{\hat{\Lambda}}$.
The character $\chi_{\hat{\Lambda}}(q;Z)$ of the representation
$R_{\hat{\Lambda}}$ is a function of the variables $q$ and
$Z=\{z_{1},\ldots,z_{p-1}\}$ which encodes the weights of the
representation or, equivalently, the eigenvalues of the
representatives of the Cartan
generators $h^{i}$, for $i=1,\ldots p-1$ of the global subalgebra
$A_{p-1}\subset\hat{A}_{p-1}$ together with those of the derivation 
$L_{0}$ acting in $\mathcal{V}_{\hat{\Lambda}}$. Explicitly we define
\begin{eqnarray}
\chi_{\hat{\Lambda}}(q;Z) & = &
{\rm
  Tr}_{\mathcal{V}_{\hat{\Lambda}}}\left[q^{-R_{\hat{\Lambda}}(L_{0})}
\prod_{j=1}^{p-1}\,  z_{j}^{R_{\hat{\Lambda}}(h^{j})}\right]
\nonumber
\end{eqnarray}
Any representation of $\hat{A}_{p-1}$ must also provide a
representation of the global subalgebra $A_{p-1}$. Thus the affine
character must have an expansion in terms of $A_{p-1}$ characters of
the form
\begin{eqnarray}
\chi_{\hat{\Lambda}}(q;Z) & = &
\sum_{\Lambda\in \mathcal{L}_{W}^{+}} \,
b_{\hat{\Lambda}}^{\Lambda}(q) \, \chi_{\Lambda}(Z)
\nonumber
\end{eqnarray}
The coefficients $b_{\hat{\Lambda}}^{\Lambda}(q)$ are polynomials in
$q$ with non-negative integral coefficients. They are known as the {\em
  branching functions} for the embedding of $A_{p-1}$ in
$\hat{A}_{p-1}$. Another way to characterise them is to pick out only
those vectors in the representation space
$\mathcal{V}_{\hat{\Lambda}}$ which are highest weight with respect to
the global generators. Thus we define, for each dominant integral weight
$\Lambda$ of $A_{p-1}$ with
$\Lambda=\sum_{i=1}^{p-1}\psi^{i}\Lambda_{(i)}$, the following subspace:
\begin{eqnarray}
\mathcal{V}_{\hat{\Lambda}}^{\Lambda} & = & \{ |\hat{\lambda}\rangle
\in \mathcal{V}_{\hat{\Lambda}}: h^{i}
| \hat{\psi}\rangle=\psi^{i}
| \hat{\psi}\rangle, e^{i}|\hat{\psi}\rangle=0 \,\,\,i=1,\ldots,p-1\}
\nn
\end{eqnarray}
Then we have
\be
b_{\hat{\Lambda}}^{\Lambda}(q)= {\rm Tr}_
{\mathcal{V}^{\Lambda}_{\hat{\Lambda}}}\left[q^{-R_{\hat{\Lambda}}(L_{0})}
\right]
\nn
\ee
\para
Remarkably a relation between the large $N=Lp+M$ limit (with fixed $M$
and $p$) of the object
$K_{\Lambda}(q)$ defined in (\ref{zhat3}) above and a particular
affine branching function of $\hat{A}_{p-1}$
is obtained in \cite{morekostka}, proving an earlier conjecture of 
\cite{kirillov}. In
particular, we must consider the integrable representation
$R_{\hat{\Lambda}}$ with $\hat{\Lambda}=k\Lambda_{(M)}$. The primary
states in the representation (i.e. those with the lowest $L_{0}$
eigenvalue) transform in the $A_{p-1}$ representation with
$\Lambda_{0}=k\Lambda_{(M)}$; this corresponds to the $k$-fold
symmetrisation of the $M^{\rm th}$ antisymmetric power of the fundamental
representation. This is indeed the expected representation \eqn{rep} for the
ground
state of the model.
This representation has congruence class
$\mathbb{P}(\Lambda_{0})=kM$ mod $p$ and the remaining dominant
integral weights $\Lambda$ for which $b^{\Lambda}_{\hat{\Lambda}}(q)$
is non-zero necessarily lie in the same congruence class. Corollary
4.8 of  \cite{morekostka} states that, for all
$\Lambda\in \mathcal{L}_{W}^{+}(kM)$, we have
\begin{eqnarray}
\lim_{N\rightarrow \infty}\,\, K_{\Lambda}(q) & = &
b^{\Lambda}_{k\Lambda_{(M)}}(q)
\label{cor}
\end{eqnarray}
This result has its origin \cite{ny1} in the relation between the Kostka
polynomials and the partition function of an integrable
$A_{p-1}$ spin chain to which we alluded above. Under favourable
conditions, the relevant 
spin chain is believed to go over to the $SU(p)$ Wess-Zumino-Witten
model in the continuum limit \cite{Aff}. 
Kostka polynomials also
appear \cite{ny2} in the partition function of the so-called {\em RSOS models}, which yield coset
conformal field theories with affine Lie algebra symmetry in the continuum
limit.     
\para
Incorporating the above limit in the reduced partition function as given in
(\ref{zhat3}) we reach
our final result
\begin{eqnarray}
\lim_{N\rightarrow \infty} \hat{\mathcal{Z}} =
\prod_{j=1}^{\infty} \frac{1}{(1-q^{j})}\,\,\,\, \chi_{k\Lambda_{(M)}}
\left(q;Z\right)
\label{final}
\end{eqnarray}
where the variables $Z=\{z_{1},\ldots,z_{p-1}\}$ are related to the
$\mathfrak{su}(p)$ fugacities of the matrix model by equations
(\ref{xtilde}) and (\ref{zx}). The prefactor encoding the excitation
spectrum of the $\mathfrak{u}(1)$ sector of the model
precisely corresponds to the partition function of a chiral boson.

\section*{Acknowledgements}

We are grateful to Sean Hartnoll, Gautam Mandal and Shiraz Minwalla for many
useful conversations.  DT and CT give thanks to the theory group in TIFR for
their very kind hospitality while this work was undertaken. DT is also grateful
to the Stanford Institute for Theoretical Physics for hospitality while this
work was written up.
We are supported by STFC and by the European Research Council under the
European Union's Seventh Framework Programme
(FP7/2007-2013), ERC grant agreement STG 279943, ``Strongly Coupled Systems".

\appendix

\section{Appendix: Proofs of Two Classical Identities}\label{idsec}

In this appendix we prove the two classical identities that we used to exhibit the existence of a Kac-Moody algebra. Assuming $m \ge n$, they are

\para
{\bf Identity 1:}
\be  
&& \varphi_i^\dagger Z^{\dagger n} Z^m \varphi_l -  \frac 1 p \displaystyle \sum_{r=0}^{m-1}\sum_{s=0}^{n-1}( \varphi_{i'}^\dagger Z^{r} Z^{\dagger s} \varphi_{i'})(\varphi_{i}^\dagger Z^{\dagger n-1-s} Z^{m-1-r}  \varphi_l) 
- \delta_{il} (\cdots) \nn \\
&& \qquad \qquad \qquad \qquad \qquad \qquad \sim \left( \frac{kN}{p} \right)^n \varphi^\dagger_i Z^{m-n} \varphi_l - \delta_{il} (\cdots) \label{id1classical}\ee
\para
{\bf Identity 2 (Classical Version):}
\be
\sum_{r=0}^{m-1} \varphi_{i'}^\dagger Z^{r} Z^{\dagger n-1} Z^{m-1-r}  \varphi_{i'} \sim p \left( \frac {(k+p)N}{p} \right)^n  \delta_{mn}
\label{id2classical}\ee
where  $\sim$ means up to $1/N$ corrections and, in the first identity,  we subtract off the $il$ trace on both sides.

\para
The phrase ``up to order $1/N$ corrections" implicitly includes a restriction on the kind of classical solutions on which we should evaluate these expressions. Roughly speaking, the solutions shouldn't deviate by ${\cal O}(N)$ from the ground state. We start by describing in more detail what this means. 

\para
For the $p=1$ matrix model, the ground state was given in \cite{alexios}
%
%
%
\be Z= Z_0(N) \equiv \sqrt{k}\left(\begin{array}{cccccc}  0\ & 1\ &  & &  \\  & 0\ & \sqrt{2} & &  \\   & & & \ddots &   \\ &  &  & 0 & \sqrt{N-1} &  \\  & & & & 0  \end{array}\right)\ \ \ {\rm and}\ \ \ \varphi = \varphi_0(N) \equiv \sqrt{k} \left(\begin{array}{c} 0 \\  0 \\ \vdots \\ 0\\ \sqrt{N}\end{array}\right)\nn\ee
together with $\alpha =  ({\omega^2}/{B})\,{\rm diag} (N-1, N-2,\ldots, 2,1,0)$.

\para
It is simple to embed these solutions in the more general matrix model. The number of ground states now depends on the relative values of $N$ mod $p$. It is simplest when $N$ is divisible by $p$. In this case there is a unique ground state which takes the block diagonal form
\be Z= Z_0(N/p) \otimes {\bf 1}_p\ \ \ ,\ \ \ \varphi = \varphi_0(N/p)\otimes {\bf 1}_p\label{ground}\ee
%
%
%
where we've written the $\varphi_i$ (with $i=1,\ldots,p$) as an $N\times p$ matrix, denoted by $\varphi$. 
%
%
%

\para
If $N$ is not divisible by $p$ then there are multiple classical ground states, transforming in the representation \eqn{rep} \cite{us2}. For example, if $N = 1$ mod $p$ then each of the blocks has $Z_0((N-1)/p)$, except for one which has $Z_0((N + p-1)/N)$. There are $p$ such choices; these ground states transform in the ${\bf p}$ of the $SU(p)$ global symmetry. 
Similarly, if $N= q$ mod $p$ then there are $p\choose q$ ground states, transforming in the $q^{\rm th}$ antisymmetric representation of $SU(p)$. 

\para
In what follows, we will assume that $N$ is divisible by $p$. 
Now we can make our statement about ${\cal O}(1/N)$ corrections more precise. We should treat $\varphi \sim O(N^{1/2})$ and $Z \sim O(N^{1/2})$, since in the ground state the largest components of either scale like the square root of $N$, and even when contracting indices there is only one non-zero entry per row or column. (This is important to check because there are $O(N)$ components, which could upset our counting.) We will evaluate the identities on states which differ from the ground state by ${\cal O}(1)$  when measured naturally by the norm squared of $\delta Z$ and $\delta \phi$. It is important that these states still satisfy the Gauss' law constraint \eqn{constraint}.

\para
These restrictions  make it fairly straightforward to prove the classical version of Identity 2. 
%
%
Consider a linear expansion of the left-hand side around the ground state in powers of $N^{1/2}$; we obtain the zeroth order term plus something we can bound by $\epsilon N^{n-1/2}$. If we decide to neglect terms of this order, we can simply substitute the expression for the ground state into the left hand side. It is trivial to check that $Z^\dagger \varphi_i = 0$, and hence the only contribution is from $\varphi_i^\dagger Z^{\dagger n-1} Z^{m-1} \varphi_i$.

\para
Next, observe that
\be Z^{\dagger m} Z^m \varphi_i = (k+p)^m \frac {(N/p-1)!}{(N/p-1-m)!} \varphi_i \sim \left( \frac {(k+p)N} p \right)^m \varphi_i \nn \ee
Upon using $Z^\dagger \varphi_i = 0$ once more, the $\delta_{nm}$ factor in \eqn{id2classical} follows. The final ingredient is to observe $\varphi_i^\dagger \varphi_i = kN$, completing the proof of \eqn{id2classical}.

\para
Identity 1 is a little harder to prove. 
Let us start by rewriting it slightly:
\be  
 \varphi_i^\dagger Z^{\dagger n} Z^{n+m} \varphi_l -  \frac 1 p \displaystyle \sum_{r=0}^{n+m-1}\sum_{s=0}^{n-1}( \varphi_{i'}^\dagger Z^{r} Z^{\dagger s} \varphi_{i'})(\varphi_{i}^\dagger Z^{\dagger n-1-s} Z^{n+m-1-r}  \varphi_l) 
 \sim' \left( \frac{(k+p)N}{p} \right)^n \varphi^\dagger_i Z^m \varphi_l  \nn\ee
where for brevity the prime $'$ denotes (asymptotic) equality of the $il$-traceless parts. We will proceed by firstly showing that only one term in the double sum contributes at leading order, namely that obtained at $r=s=0$, reducing the problem to proving
\be \varphi_i^\dagger Z^{\dagger n} Z^{n+m} \varphi_l -  \frac {kN} p (\varphi_{i}^\dagger Z^{\dagger n-1} Z^{n+m-1}  \varphi_l) \  \sim' \ \left( \frac{(k+p)N}{p} \right)^n \varphi^\dagger_i Z^m \varphi_l \nn \ee
Then we will inductively demonstrate that
\be \varphi_i^\dagger Z^{\dagger n} Z^{n+m} \varphi_l \sim' (n+1) \left( \frac{(k+p)N}{p} \right)^n \varphi^\dagger_i Z^m \varphi_l \label{id2classicalind} \ee
from which the original identity follows immediately.

\para
So to begin, let us estimate the size of the terms we wish to keep. The traceless part of the right-hand side vanishes in the ground state, so we must sacrifice at least one term for something of order $\epsilon$; this is then generically non-vanishing. Therefore, the right-hand side is of order $O(\epsilon N^{n+(m+1)/2})$.

\para
We can now consider a single term of the double sum at general $(r,s)$. The traceless part of the second bracket, $(\varphi_{i}^\dagger Z^{\dagger n-1-s} Z^{n+m-1-r}  \varphi_l)$, vanishes in the ground state, and hence is at most order $O(\epsilon N^{n+(m-r-s-1)/2})$. Thus the first bracket must be at least of order $O(N^{1+(r+s)/2})$. But this $N$-scaling is only possible if all terms in the first bracket come from the ground state, when this term vanishes by the observations above -- except for $r=s=0$.

\para
This leaves us only with deriving \eqn{id2classicalind}. We will induct on $n$ to establish this; note that the case $n=0$ is trivial. Write
\be \varphi_i^\dagger Z^{\dagger n} Z^{n+m} \varphi_l =' \varphi_i^\dagger Z^m Z^{\dagger n} Z^{n} \varphi_l - \varphi_i^\dagger [Z^m, Z^{\dagger n}] Z^{n} \varphi_l \nn \ee
The first term is simple to handle. Since at leading order $\varphi_i^\dagger Z^m = 0$, we can safely make the approximation $Z^{\dagger n} Z^n \varphi_i \sim ((k+p)N/p)^n \varphi_i$.

\para
The second term can be expanded into a double sum, and simplified slightly using the asymptotic version of Gauss' constraint, $[Z,Z^\dagger] \sim - \varphi \varphi^\dagger$. Then almost all terms can be shown to be subleading, using the ideas above, except for the one where we have the $\varphi \varphi^\dagger$ appearing at the far left. Hence
\be \varphi_i^\dagger Z^{\dagger n} Z^{n+m} \varphi_l &\sim'& \left(\frac {(k+p)N}{p}\right)^n \varphi_i^\dagger Z^m \varphi_l + \varphi_i^\dagger \varphi_j \varphi_j^\dagger Z^{\dagger n-1} Z^{m-1} Z^{n} \varphi_l \nn \\
&\sim'& \left(\frac {(k+p)N}{p}\right)^n \varphi_i^\dagger Z^m \varphi_l +\left(\frac {(k+p)N}{p}\right) \varphi_i^\dagger Z^{\dagger n-1} Z^{n-1+m} \varphi_l \nn \ee
where we have also used the trick of separating the $ij$ and $jl$ traces out, discarding more irrelevant terms. Finally, applying the inductive hypothesis to the second term, we establish \eqn{id2classicalind}, and hence identity 1.

\section{Appendix: Kostka Polynomials}\label{kostkasec}

\para

In this appendix we give an explicit description of the Kostka
polynomial $K_{\lambda, \mu}(q)$ due to Kirillov and Reshetikhin \cite{Kir}.

\para

Given $\lambda$, $\mu\in \mathcal{P}$, we define a sequence of
partitions $\nu^{(K)}$ with $K=0,1,2,\ldots,\ell(\lambda)-1$ with
$\nu^{(0)}=\mu$ and
\be
| \nu^{(K)}| = \sum_{j\geq K+1} \, \lambda_{j}
\label{c1}
\ee
For each such sequence we define the {\em vacancy numbers}
\begin{eqnarray}
\mathbb{P}^{(K)}_{n} &  = &
\sum_{j\geq 1} \left[ {\rm min}\{n,\nu^{(K+1)}_{j}\} -2 {\rm min}\{n,\nu^{(K)}_{j}\}
+{\rm min}\{n,\nu^{(K-1)}_{j}\} \right]
\nonumber
\end{eqnarray}
for all positive integers $n$ and  $K=0,1,2,\ldots,\ell(\lambda)-1$
with the understanding that $\nu^{(\ell(\lambda))}\equiv 0$.  An {\em
admissible configuration}, $\{\nu\}$ is any such sequence of
partitions with non-negative
vacancy numbers, i.e.
\be
 \mathbb{P}^{(K)}_{n}\geq 0
\nn
\ee
for all values of $n$ and $K$. The {\em charge} $c(\{\nu\})$ of an
admissible configuration is defined as
\begin{eqnarray}
c\left(\{\nu\}\right) & = & n[\mu]\,+\, \sum_{K=1}^{\ell(\lambda)-1}
\,
\left( \mathbb{M}\left[\nu^{(K)},\nu^{(K)}\right]\,-\,
\mathbb{M}\left[\nu^{(K)},\nu^{(K-1)}\right]\right)
\nn
\end{eqnarray}
where, for any two partitions $\rho$, $\kappa\in\mathcal{P}$, we define
the function
$\mathbb{M}: \mathcal{P}\times \mathcal{P}\rightarrow \mathbb{Z}_{\geq
  0}$ by
\begin{eqnarray}
\mathbb{M}\left[\rho,\kappa\right] & = & \sum_{i,j\geq 1}\,
{\rm min}\{\rho_{i},\kappa_{j}\} \nn
\end{eqnarray}
\para
Finally the Kostka polynomial can be defined as a sum over all
admissible configurations; explicitly,
\begin{eqnarray}
K_{\lambda,\mu}(q) & = & \sum_{\{\nu\}}\,q^{c\left[\{\nu\}\right]}\,\,
\prod_{K=1}^{\ell(\lambda)-1}\, \prod_{n \geq 1} \left
  [ \begin{array}{c} \mathbb{P}^{(K)}_{n}+m_{n}\left(\nu^{(K)}\right)
    \\ m_{n}\left(\nu^{(K)}\right) \end{array} \right]_{q}
\nonumber
\end{eqnarray}
where we define the $q$-binomial coefficient
\begin{eqnarray}
\left[ \begin{array}{c} m \\ n \end{array} \right]_{q} & = &
\frac{\varphi_{m}(q)}{\varphi_{n}(q)\varphi_{m-n}(q)}
\nonumber
\end{eqnarray}
and, as in the text, we have
$\varphi_{n}(q)=\prod_{j=1}^{n}(1-q^{j})$.

\subsubsection*{Ground State Energy}

As in the text we set $\mu=(k^{N})$ and look for the ground state by
searching for the partition $\lambda$, satisfying $|\lambda|=kN$ and
$\ell(\lambda)\leq p$, such that $K_{\lambda,
  (k)^{N}}(q)$ yields the leading term in the $q$ expansion for
$|q|\ll1$. The general formula given above can be simplified \cite{Kir} in the case $k=1$
where $\mu=(1^{N})$. We find
\begin{eqnarray}
 K_{\lambda,(1^{N})}(q) & = & q^{n\left[\lambda^{T} \right]} \,
\frac{\prod_{j=1}^{N} {(1-q^{j})}}{H(q)}
\nonumber
\end{eqnarray}
where $H(q)$ is the hook-length polynomial given by
\begin{eqnarray}
H(q) & = & \prod_{x\in \mathbb{Y}(\lambda)}\, \left(1- q^{h(x)}\right)
\nonumber
\end{eqnarray}
Here the product is over the boxes $x=(r,s)$ of the Young diagram
$\mathbb{Y}(\lambda)$ corresponding to the partition $\lambda$ and
$h(x)=\lambda_{r}+\lambda^{T}_{s}-r-s+1>0$ is the length of the hook
passing through box $x$.
\para
As explained in the text, the minumum for the case $k=1$ is
attained for the partition $\lambda_{0}=((L+1)^{M},L^{p-M})$ where
$N=Lp+M$ for non-negative integers $L$ and $M<p$ which gives a
ground state energy
\be
E_{0}(1,p,N)=\frac{1}{2}L(L-1)p+LM
\nn
\ee
It is instructive to reproduce this result from the general formula
given above in the case $\mu=(1^{N})$ and
$\lambda=\lambda_{0}$. According to the recipe we must
find a sequence of partitions $\nu^{(K)}$ for $K=0,1,2,\ldots,
\ell(\lambda_{0})-1=p-1$ with $\nu^{(0)}=\mu=(1^{N})$ and
\begin{eqnarray}
| \nu^{(K)}| = \sum_{j\geq K+1} \, (\lambda_{0})_{j}  & = &
 \left\{\begin{array}{lc}
(M-K)(L+1)+L(P-M) \ \ \ \ \
& 0\leq K \leq M \\
L(p-K)  & M+1 \leq K \leq p-1 \end{array}\right.
\nn
\end{eqnarray}
%
%
%
with non-negative occupation numbers which minimises the charge
$c(\{\nu\})$. It not hard to see that this is achieved by maximising
the number of parts in each partition $\nu^{(K)}$. Thus we set
\begin{eqnarray}
\nu^{(K)}  = \left\{\begin{array}{cl}
\left( 1^{\left((M-K)(L+1)+L(P-M)\right)}\right)
\ \ \ \ &  0\leq K \leq M \\
\left( 1^{\left(L(p-K)\right)}\right)  & M+1 \leq K \leq p-1
\end{array}\right.
\nn
\end{eqnarray}
%
One may then check that the corresponding occupation numbers are
non-negative and that
\begin{eqnarray}
c\left(\{\nu\}\right) & = & E_{0}(1,p,N) =\frac{1}{2}L(L-1)p+LM
\nn
\end{eqnarray}
The above configuration has a straightforward generalisation to $k\geq
1$. As in the text we set
\be
\lambda_{0}=((kL+k)^{M},(kL)^{p-M})
\nn
\ee
and one simply scales each partition $\nu^{(K)}$
in the configuration by a factor of $k$ setting
\begin{eqnarray}
\nu^{(K)} = \left\{\begin{array}{cl}
\left( k^{\left((M-K)(L+1)+L(P-M)\right)}\right)
\ \ \ \ \  & 0\leq K \leq M \\
\left( k^{\left(L(p-K)\right)}\right)  &  M+1 \leq K \leq p-1
\end{array}\right.
\nn\end{eqnarray}
%
%
The vacancy numbers are remain non-negative and the
charge of the configuration scales
linearly with $k$. Thus the new ground state energy is
\begin{eqnarray}
c\left(\{\nu\}\right) & = & E_{0}(k,p,N) =\frac{k}{2}L(L-1)p+kLM
\nn
\end{eqnarray}
which is the result stated in the main text.

\section{Appendix: (Affine) Lie Algebra Conventions}\label{liesec}

Here we give our conventions for the simple Lie algebra
$A_{p-1}=\mathfrak{sl}(p,\mathbb{C})$ and its Affine counterpart
$\hat{A}_{p-1}$.
\paragraph{}
For $A_{p-1}$, we work in a Chevalley basis with generators
$\{h^{i},e^{i},f^{i}; i=1,\ldots p-1\}$ with brackets
\be
[h^{i},h^{j}]  =  0 \ \ \ ,\ \ \
[h^{i},e^{j}] = A_{ji}e^{j} \ \ \ ,\ \ \
[h^{i},f^{j}]  =-A_{ji}f^{j} \ \ \ ,\ \ \
[e^{i},f^{j}]  =  \delta_{ij}h^{i} \nonumber
\ee
where $A_{ij}$ is the $A_{p-1}$ Cartan matrix. Weights of each irreducible
representation lie in the weight lattice
\begin{eqnarray}
\mathcal{L}_{W} & = & {\rm Span}_{\mathbb{Z}}\{ \Lambda_{(i)},
i=1,\ldots, p-1\}
\nn
\end{eqnarray}
whose basis vectors are the fundamental weights
$\Lambda_{(i)}$. Finite-dimensional, irreducible representations
$R_{\Lambda}$ are labelled by a highest weight $\Lambda$, lying in the
positive weight lattice
 \begin{eqnarray}
\mathcal{L}^{+}_{W} & = & {\rm Span}_{\mathbb{Z}_{\geq 0}}\{ \Lambda_{(i)},
i=1,\ldots, p-1\}
\nn
\end{eqnarray}
We denote the corresponding representation space
$\mathcal{V}_{\Lambda}$. For each weight of $R_{\Lambda}$ there is
an element
\begin{eqnarray}
\Psi = \sum_{i=1}^{p-1} \psi^{i}\Lambda_{(i)}
\end{eqnarray}
of $\mathcal{L}_{W}$ with Dynkin labels $\psi^{i}\in \mathbb{Z}$. We then have a
basis vector $|\Psi\rangle$
of $\mathcal{V}_{\Lambda}$ which is a simultaneous
eigenvector of the Cartan
generators satisfying
\begin{eqnarray}
R_{\Lambda}\left( h^{i}\right)|\Psi\rangle & = & \psi^{i}|\Psi\rangle
\nn
\end{eqnarray}
for $i=1,\ldots, p-1$.
\paragraph{}
For the affine Lie algebra $\hat{A}_{p-1}$ we have Chevalley
generators
$\{h^{i},e^{i},f^{i}; i=0,\ldots p-1\}$ with brackets
\be
[h^{i},h^{j}] = 0 \ \ \ ,\ \ \
[h^{i},e^{j}]  =  \hat{A}_{ji}e^{j} \ \ \ ,\ \ \
[h^{i},f^{j}]  = -\hat{A}_{ji}e^{j} \ \ \ ,\ \ \
[e^{i},f^{j}] =  \delta_{ij}h^{i}
\nn
\ee
where $\hat{A}_{ij}$ is the affine Cartan matrix. The basis elements
with $i>0$ generate an $A_{p-1}$ subalgebra. Weights of an integrable 
representation have an expansion
\begin{eqnarray}
\hat{\Psi} & = & \sum_{i=0}^{p-1} \hat{\psi}^{i}\hat{\Lambda}_{(i)} + n \delta
\label{exp}
\end{eqnarray}
for integers $\hat{\psi}_{i}$ and $n$ 
where $\delta$ is the imaginary root. The fundamental weights of
$\hat{A}_{p-1}$ can be written as
\begin{eqnarray}
\hat{\Lambda}_{(i)} & = & \hat{\Lambda}_{(0)}+{\Lambda}_{(i)}
\nn
\end{eqnarray}
for $i>0$, where $\Lambda_{(i)}$ are fundamental weights of the global
$A_{p-1}$ subalgebra.
\paragraph{}
The integrable representations $R_{\hat{\Lambda}}$
of $\hat{A}_{p-1}$ are characterised by
a highest weight $\hat{\Lambda}$ with non-negative Dynkin labels, and have
the representation space
$\mathcal{V}_{\hat{\Lambda}}$. Each weight of $R_{\hat{\Lambda}}$ 
has an expansion of the form (\ref{exp}). The
corresponding basis vector $|\hat{\Psi}\rangle$
of $\mathcal{V}_{\hat{\Lambda}}$ is a simultaneous
eigenvector of the Cartan
generators, with
\begin{eqnarray}
 R_{\hat{\Lambda}}\left(h^{i}\right)
|\hat{\Psi}\rangle & = & \hat{\psi}^{i}|\hat{\Psi}\rangle
\nn
\end{eqnarray}
for $i=1,\ldots, p-1$, and the derivation or grading operator, with
\begin{eqnarray}
-R_{\hat{\Lambda}}\left(L_{0}\right)|\hat{\Psi}\rangle & = &  n|\hat{\Psi}\rangle
\nn
\end{eqnarray}

\end{document}